\documentclass[12pt]{article}

\usepackage{geometry}
\usepackage{amssymb,latexsym,amsmath}
\usepackage{mathrsfs}
\usepackage{graphicx}
\usepackage{color}
\usepackage{bbm} 

\numberwithin{equation}{section}

\newcommand{\be}{\begin{equation}}
\newcommand{\ee}{\end{equation}}
\newcommand{\bea}{\begin{eqnarray}}
\newcommand{\eea}{\end{eqnarray}}
\newcommand{\nn}{\nonumber}
\newcommand{\dd}{\textrm{d}}
\newcommand\diff{\mathrm{d}}

\newcommand{\ti}{\widetilde}
\newcommand{\C}{\mathbb{C}}

\newcommand{\Fo}{{\mathcal{F}}} 
\newcommand{\Go}{{\mathcal{G}}}
\newcommand{\Ho}{{\mathcal{H}}}

\newcommand{\const}{\kappa} 

\begin{document}

\begin{titlepage}

\begin{center}

\today

\vskip 2.3cm 

{\Large \bf Comments on supersymmetric solutions of}

\vskip 4mm

{\Large \bf minimal gauged supergravity in five dimensions}

\vskip 15mm

{Davide Cassani${}^{\,a}$, Jakob Lorenzen${}^{\,b}$, and Dario Martelli${}^{\,b}$}

\vskip 1cm

${}^{\,a}$ \textit{Sorbonne Universit\'es UPMC Paris 06,\\ [1mm] 
UMR 7589, LPTHE, F-75005, Paris, France\\ [1mm]
and\\ [1mm]
CNRS, UMR 7589, LPTHE, F-75005, Paris, France}

\vskip 2mm

{\small \texttt{davide.cassani AT lpthe.jussieu.fr}}

\vskip 6mm

${}^b$\textit{ Department of Mathematics, King's College London, \\ [1mm]
The Strand, London WC2R 2LS,  United Kingdom\\}

\vskip 2mm

{\small \texttt{jakob.lorenzen AT kcl.ac.uk, dario.martelli AT kcl.ac.uk}}

\end{center}

\vskip 2 cm

\begin{abstract}
\noindent  
We investigate supersymmetric solutions of minimal gauged supergravity in five dimensions,  in the timelike class. We propose an ansatz based on a four-dimensional local \emph{orthotoric} K\"ahler metric  and reduce the problem to a single sixth-order equation for two functions, each of one variable. We find an analytic, asymptotically locally AdS solution comprising five parameters. For a conformally flat boundary, this reduces to a previously known solution with three parameters, representing the most general solution of this type known in the minimal theory. We discuss the possible relevance of certain topological solitons contained in the latter to account for the supersymmetric Casimir energy of dual superconformal field theories on $S^3\times \mathbb R$. Although we obtain a negative response, our analysis clarifies several aspects of these solutions. In particular, we show that there exists a unique regular topological soliton in this family. 

\end{abstract}

\end{titlepage}

\pagestyle{plain}
\setcounter{page}{1}
\newcounter{bean}
\baselineskip18pt

\tableofcontents

\section{Introduction}

Supersymmetric solutions to five-dimensional supergravity play an important role in the AdS/CFT correspondence. In particular, solutions to minimal gauged supergravity describe universal features of four-dimensional $\mathcal N=1$ superconformal field theories (SCFT's).
The boundary values of the fields sitting in the gravity multiplet, that is the metric, the graviphoton and the gravitino, are interpreted on the field theory side as background fields coupling to the components of the energy-momentum tensor multiplet, namely the energy-momentum tensor itself, the $R$-current and the supercurrent.
When the solution is asymptotically Anti de Sitter (AAdS), and one chooses to work in global coordinates, the dual SCFT is defined on the conformally flat boundary $S^3 \times \mathbb R$. Solutions that are only asymptotically {\it locally} Anti de Sitter (AlAdS) describe SCFT's on non conformally flat boundaries. For instance, they can describe SCFT's on backgrounds comprising a squashed $S^3$.

The conditions for obtaining a supersymmetric solution to minimal gauged supergravity in five dimensions were presented about a decade ago in~\cite{Gauntlett:2003fk}. These fall in two distinct classes, timelike and null. The timelike class is somewhat simpler and takes a canonical form determined by a four-dimensional K\"ahler base.
 Using this formalism, in~\cite{Gutowski:2004ez} the first example of supersymmetric AAdS black hole free of closed timelike curves was constructed. Other AAdS solutions were obtained by different methods in~\cite{London:1995ib,Klemm:2000vn,Cvetic:2004hs,Chong:2005hr}, with the solution of \cite{Chong:2005hr} being the most general in that it encompasses the others as special sub-cases.
The solution of~\cite{Chong:2005hr} also includes the most general AAdS black hole known in minimal gauged supergravity; in the supersymmetric limit, this was shown to take timelike canonical form in~\cite{Kunduri:2006ek}.
The formalism of~\cite{Gutowski:2004ez} also led to construct AlAdS solutions, see~\cite{Gauntlett:2004cm,Figueras:2006xx,Cassani:2014zwa} for the timelike class and \cite{Bernamonti:2007bu} (based on \cite{Klemm:2000nj}) for the null class. 
 
In this paper, we come back to the problem of finding supersymmetric solutions to minimal gauged supergravity. Our motivation is twofold: on the one hand we would like to investigate the existence of black holes more general than the one of \cite{Chong:2005hr}.   Indeed, in the supersymmetric limit the black hole of \cite{Chong:2005hr} has two free parameters, while one may expect the existence of a three-parameter black hole, whose mass, electric charge, and two angular momenta are constrained only by the BPS condition (see e.g.~\cite{Kunduri:2006uh} for a discussion).
Our second motivation is to construct supersymmetric A(l)AdS solutions with no horizon, that are of potential interest as holographic duals of pure states of SCFT's in curved space.
 For instance, large $N$ SCFT's on a squashed $S^3\times \mathbb R$, where the squashing only preserves a $U(1)\times U(1)$ symmetry, should be described by an AlAdS supergravity solution that is yet to be found.  It is reasonable to assume that this would preserve $U(1)\times U(1)\times \mathbb R$ symmetry also in the bulk, so it would carry mass and two angular momenta in addition to the electric charge.

Within the approach of~\cite{Gauntlett:2003fk}, we introduce an ansatz for solutions in the timelike class based on an {\it orthotoric} K\"ahler metric. This has two commuting isometries and depends on two functions, each of one variable. We reduce the problem of obtaining supersymmetric solutions to a single equation of the sixth order for the two functions. This follows from a constraint on the conditions of~\cite{Gauntlett:2003fk},  that we formalise in full generality.
We find a polynomial solution to the sixth-order equation depending on three non-trivial parameters. Subsequently, in the process of constructing the rest of the five-dimensional metric we obtain two additional parameters, leading to an AlAdS solution. 
We show that when these two extra parameters are set to zero, our solution is AAdS and is related to that of~\cite{Chong:2005hr} by a change of coordinates. We observe that for specific values of the parameters of~\cite{Chong:2005hr} the change of coordinates becomes singular, and interpret this in terms of a scaling limit of the orthotoric ansatz, leading to certain non-orthotoric K\"ahler metrics previously employed in the search for supergravity solutions. 
This proves that our orthotoric ansatz, together with its scaling limit, encompasses \emph{all} known supersymmetric solutions to minimal gauged supergravity belonging to the timelike class.

After having completed the general study, we focus on certain non-trivial geometries with no horizon contained in the solution of~\cite{Chong:2005hr}, called {\it topological solitons}. These are a priori natural candidates to describe pure states of dual SCFT's. We report on an attempt to match holographically the vacuum state of an $\mathcal N=1$ SCFT on the cylinder $S^3\times \mathbb R$, and in particular the non-vanishing supersymmetric vacuum expectation values of the energy and $R$-charge~\cite{Assel:2015nca}. Some basic requirements following from the supersymmetry algebra lead us to consider a 1/2 BPS topological soliton presented in~\cite{Cvetic:2005zi}. Although a direct comparison of the charges shows that this fails to describe the vacuum state of the dual SCFT, in the process we clarify several aspects of such solution. We also show that the 1/4 BPS topological solitons with an $S^3\times \mathbb R$ boundary contain a conical singularity.

The structure of the paper is as follows. In section~\ref{GGclassification} we review the equations of~\cite{Gauntlett:2003fk} for supersymmetric solutions of the timelike class and obtain the general form of the sixth-order constraint on the K\"ahler metric. 
 In section~\ref{sec:orthotoric} we present our orthotoric ansatz and find a solution to the constraint; then we construct the full five-dimensional solution and relate it to that of~\cite{Chong:2005hr}. The scaling limit of the orthotoric metric is presented in section~\ref{sec:scalinglimit}. In section~\ref{sec:toposoli} we make the comparison with the supersymmetric vacuum of an SCFT on $S^3\times \mathbb R$, and discuss further aspects of the topological soliton solutions. In section~\ref{sec:conclusions} we draw our conclusions. In appendix~\ref{SO4appendix} we prove  uniqueness of a supersymmetric solution  within an ansatz with $SO(4)\times \mathbb R$ symmetry, while in appendix~\ref{app:uplift} we discuss the obstructions to uplift the 1/2 BPS topological solitons of~\cite{Cvetic:2005zi} to type IIB supergravity on Sasaki-Einstein manifolds and make some comments on the dual field theories.

\section{Supersymmetric solutions from K\"ahler bases}\label{GGclassification}

In this section we briefly review the conditions for bosonic supersymmetric solutions to minimal five-dimensional gauged supergravity found in~\cite{Gauntlett:2003fk}, focusing on the timelike class. We also provide a general expression of a constraint first pointed out in an example in~\cite{Figueras:2006xx}.

The bosonic action of minimal gauged supergravity is\footnote{We work in ($-$++++) signature. Our Riemann tensor is defined 
as $R^\mu{}_{\nu\kappa\lambda} = \partial_\kappa \Gamma^\mu_{\nu\lambda} + \Gamma^\mu_{\kappa\sigma}\Gamma^\sigma_{\nu\lambda} - \kappa \leftrightarrow \lambda$,
and the Ricci tensor is $R_{\mu\nu}= R^\lambda{}_{\mu\lambda\nu}$.}
\be\label{5Daction}
S \ = \ \frac{1}{2\kappa_5^2}\int \left[ (R_{5} + 12g^2) *_{5}\!1 - \frac{1}{2} F\wedge *_{5} F + \frac{1}{3\sqrt3}  A\wedge F\wedge F \right]\ ,
\ee
where $R_{5}$ is the Ricci scalar of the five-dimensional metric $g_{\mu\nu} $, $ A$ is the  graviphoton $U(1)$ gauge field, $F = \diff A$ is its field strength, $g>0$ parameterises the cosmological constant, and $\kappa_5$ is the gravitational coupling constant. The Einstein and Maxwell equations of motion are
\bea\label{5Deom}
R^{(5)}_{\mu\nu} + 4g^2 g_{\mu\nu} - \frac{1}{2} F_{\mu\kappa}F_{\nu}{}^{\kappa} + \frac{1}{12} g_{\mu\nu} F_{\kappa\lambda}F^{\kappa\lambda} &=& 0\ ,\nn \\ [2mm]
\diff *_{5} F - \frac{1}{\sqrt 3} F \wedge F &=& 0\ .
\eea

A bosonic background is supersymmetric if there is a non-zero Dirac spinor $\epsilon$ satisfying
\be\label{KillingSpEqDirac}
\left[ \nabla^{(5)}_\mu - \frac{i}{8\sqrt 3}\left( \gamma_{\mu}{}^{\nu\kappa} - 4\delta^\nu_\mu\gamma^\kappa\right) F_{\nu\kappa} - \frac{g}{2}  \big( \gamma_\mu + \sqrt 3\,i A_\mu \big)\right]\epsilon  \ =\ 0\ ,
\ee
where the gamma-matrices obey the Clifford algebra $\{\gamma_\mu,\gamma_\nu\}=2g_{\mu\nu}$. 
Bosonic supersymmetric solutions were classified (locally) in \cite{Gauntlett:2003fk} by analysing the bilinears in $\epsilon$. It was shown that all such solutions admit a Killing vector field $V$ that is either timelike, or null. In this paper we do not discuss the null case and focus on the timelike class.

Choosing adapted coordinates such that $V = \partial/\partial y$, the five-dimensional metric can be put in the form
\be\label{5dMetricTimelike}
\diff s^2_{5} \,=\, - f^2 \left( \diff y + \omega \right)^2 + f^{-1}\, \diff s^2_B\ ,
\ee
where $\diff s^2_B$ denotes the metric on a four-dimensional base $B$ transverse to $V$, while $f$ and $\omega$ are a positive function and a one-form on $B$, respectively. 
Supersymmetry requires $B$ to be K\"ahler. This means that $B$ admits a real non-degenerate two-form~$X^1$ that is closed, {\it i.e.}\ $\diff X^1 = 0$, and such that $X^1{}_m{}^n$ is an integrable complex structure ($m,n=1,\ldots,4$ denote curved indices on $B$, and we raise the index of $X^1_{mn}$ with the inverse metric on $B$). 
It will be useful to recall that a four-dimensional K\"ahler manifold also admits a complex two-form $\Omega$ of type $(2,0)$ satisfying
\be\label{diffOmega}
\nabla_m \Omega_{np} + i P_m \Omega_{np} \,=\, 0 \ ,
\ee 
where $P$ is a potential for the Ricci form, {\it i.e.}\ $\mathcal R = \diff P$. The Ricci form is a closed two-form defined as $\mathcal R_{mn}= \frac{1}{2}R_{mnpq}(X^1)^{pq}$, where $R_{mnpq}$ is the Riemann tensor on $B$. 
Moreover, splitting $\Omega = X^2 + i X^3$, the triple of real two-forms $X^I$, $I=1,2,3$, satisfies the quaternion algebra:
\be\label{Xalgebra}
X^{I}{}_m{}^p X^{J}{}_p{}^n \, = \, -\delta^{IJ} \delta_m{}^n + \epsilon^{IJK} X^{K}{}_m{}^n\ .\ee
We choose the orientation on $B$ by fixing the volume form as ${\rm vol}_B = -\frac 12 X^1 \wedge X^1$. 
It follows that the $X^I$ are a basis of anti-self-dual forms on~$B$, {\it i.e.} $*_BX^I=-X^I$.

The geometry of the K\"ahler base $B$ determines the whole solution, namely $f$, $\omega$ in the five-dimensional metric~\eqref{5dMetricTimelike} and the graviphoton field strength $F$. The function $f$ is fixed by supersymmetry as
\be\label{solf}
f \,=\, - \frac{24g^2}{R}\ ,
\ee
where $R$ is the Ricci scalar of $\diff s^2_B$; this is required to be everywhere non-zero. 

The expression for the Maxwell field strength is
\be
F \, =\, -\sqrt 3\, \diff \Big[ f(\diff y + \omega) +\frac{1}{3g}P \Big] \ .\label{susygaugefield}
\ee
Note that the Killing vector $V$ also preserves $F$, hence it is a symmetry of the solution.

It remains to compute the one-form $\omega$. This is done by solving the equation
\be\label{G+plusG-}
\diff \omega  \, =\, f^{-1}(G^+ + G^-)\ ,
\ee
where the two-forms $G^\pm$, satisfying the (anti)-self-duality relations $*_B G^\pm = \pm G^\pm$, are determined as follows. Supersymmetry states that $G^+$ is proportional to the traceless Ricci form of $B$:\footnote{The traceless Ricci form $\mathcal R_0 = \mathcal R - \frac{R}{4} X^1$ is the primitive part of $\mathcal R$.
It has self-duality opposite to the K\"ahler form.
}
\be\label{solG+}
G^+\, =\, - \frac{1}{2g} \Big( \mathcal R - \frac{R}{4} X^1  \Big)\ .
\ee
Expanding $G^-$ in the basis of anti-self-dual two-forms as\footnote{Our $\lambda^I$ are rescaled by a factor of $2gR$ compared to those in \cite{Gauntlett:2003fk}.}
\be
G^-\, =\, \frac{1}{2gR} (\lambda^1 X^1 + \lambda^2 X^2 + \lambda^3 X^3)\ ,
\ee 
one finds that the Maxwell equation fixes $\lambda^1$ as
\be\label{sollambda1}
\lambda^1\ =\  \frac{1}{2} \nabla^2 R +\frac{2}{3} R_{mn} R^{mn} - \frac{1}{3}R^2 \ ,
\ee
where $\nabla^2$ is the Laplacian on $B$.
The remaining two components, $\lambda^2,\lambda^3$, only have to be compatible with the requirement that the right hand side of \eqref{G+plusG-} be closed,
\be\label{G+plusG-closed}
\diff \left[ f^{-1}(G^+ + G^-)\right] \ =\ 0\ .
\ee
Plugging \eqref{solf}, \eqref{solG+}, \eqref{sollambda1} in and taking the Hodge dual, one arrives at the equation
\be\label{eqforlambda23real}
{\rm Im}\left[\, \overline \Omega_m{}^n(\partial_n + iP_n)(\lambda^2+i\lambda^3)\right] + \Xi_m \ = \ 0 \ ,
\ee
where
\be
\Xi_m \ = \ R_{mn}\partial^n R + \partial_m\left( \frac 12 \nabla^2 R + \frac 23 R_{pq}R^{pq} - \frac 13 R^2 \right) \ .
\ee
Acting on \eqref{eqforlambda23real} with $\Pi_p{}^q X^3{}_q{}^m$, where $\Pi = \frac{1}{2}\left(\mathbbm{1} + i X^1 \right)$ is the projector on the $(1,0)$ part, one obtains the equivalent form
\be\label{eqforlambda23complex}
D^{(1,0)} (\lambda^2 + i \lambda^3)  +\Theta^{(1,0)} \ = \ 0 \ ,
\ee
where $D^{(1,0)}_m = \Pi_m{}^n (\nabla_n+iP_n)$ is the holomorphic K\"ahler covariant derivative, and we defined $\Theta^{(1,0)}_m = \Pi_m{}^n X^3{}_n{}^p \Xi_p$.
Eq.~\eqref{eqforlambda23complex} determines $\lambda^2 + i\lambda^3$, and hence $G^-$, up to an anti-holomorphic function.
This concludes the analysis of the timelike case as presented in~\cite{Gauntlett:2003fk}.

It was first pointed out in~\cite{Figueras:2006xx} that for equation \eqref{G+plusG-closed} to admit a solution, a constraint on the K\"ahler geometry must be satisfied. Hence not all four-dimensional K\"ahler bases give rise to supersymmetric solutions. While in~\cite{Figueras:2006xx} this was shown for a specific family of K\"ahler bases, here we provide a general formulation. 
Taking the divergence of~\eqref{eqforlambda23real} and using \eqref{diffOmega} we find
\be
\nabla^m \Xi_m \ = \ 0\ ,
\ee
that is
\be\label{constraintfinal}
\nabla^2\left( \frac 12 \nabla^2 R + \frac 23 R_{pq}R^{pq} - \frac 13 R^2 \right) + \nabla^m(R_{mn}\partial^n R) \ =\ 0\ .
\ee
We thus obtain a rather complicated sixth-order equation constraining the K\"ahler metric.\footnote{It can also be derived starting from the observation that since $D^{(1,0)}$ is a good differential, namely $(D^{(1,0)} \big )^2 =0$, equation \eqref{eqforlambda23complex} has the integrability condition \hbox{$D^{(1,0)}\Theta^{(1,0)} = 0$. } The latter is an a priori complex equation, however one finds that the real part is automatically satisfied while the imaginary part is equivalent to~\eqref{constraintfinal}.} To the best of our knowledge, this has not appeared in the physical or mathematical literature before.
 We observe that the term $(\nabla^2)^2 R + 2 \nabla^m(R_{mn}\partial^n R)$ corresponds to the real part of the Lichnerowicz operator acting on $R$, which  vanishes for \emph{extremal} K\"ahler metrics (see {\it e.g.}~\cite[sect.$\:$4.1]{ExtremalKahler}). 
 Thus in this case~\eqref{constraintfinal} reduces to $\nabla^2\left( 2 R_{pq}R^{pq} - R^2 \right) =0$. If the K\"ahler metric has constant Ricci scalar, the constraint simplifies further to $\nabla^2 (R_{pq}R^{pq})=0$.
  Finally, if the K\"ahler metric is homogeneous, or Einstein, then $\Xi = 0$ and the constraint is trivially satisfied.\footnote{Constraints on (six-dimensional and eight-dimensional) K\"ahler metrics involving higher-derivative curvature terms were also found in the study of AdS$_3$ and AdS$_2$ supersymmetric solutions to type IIB and 11-dimensional supergravity, respectively \cite{Kim:2005ez,Kim:2006qu,Gauntlett:2006ns}.}

To summarise, the five-dimensional metric and the gauge field strength are determined by the four-dimensional K\"ahler geometry up to an anti-holomorphic function. The K\"ahler metric is constrained by the sixth-order equation~\eqref{constraintfinal}. Moreover, one needs $R\neq 0$.
The conditions spelled out above are necessary and sufficient for obtaining a supersymmetric solution of the timelike class. The solutions preserve at least 1/4 of the supersymmetry, namely two real supercharges.

\section{Orthotoric solutions}\label{sec:orthotoric}

\subsection{The ansatz}

In this section we construct supersymmetric solutions following the procedure described above. We start from a very general ansatz for the four-dimensional base, given by a class of local K\"ahler metrics known as {\it orthotoric}. These were introduced in ref.~\cite{ACGweakly}, to which we refer for an account of their mathematical properties.\footnote{This ansatz was also considered in~\cite{Acharya:2006as}, however only the case $\Fo( x)=-\Go(x)$, where these are cubic polynomials, was
discussed there. In this case the metric~\eqref{orthometric} is equivalent
to the Bergmann metric on $SU(2,1)/S(U(2)\times U(1))$.
Orthotoric metrics also appear in Sasaki-Einstein geometry: as shown in~\cite{Martelli:2005wy}, the K\"ahler-Einstein bases of $L^{p,q,r}$ Sasaki-Einstein manifolds~\cite{Cvetic:2005ft} are of this type.
}  

The orthotoric K\"ahler metric reads
\be\label{orthometric}
g^2\,\diff s^2_B\ =\ \frac{\eta-\xi}{\Fo(\xi)}\diff \xi^2  + \frac{\Fo(\xi)}{\eta-\xi}(\diff \Phi + \eta \diff \Psi)^2 + \frac{\eta-\xi}{\Go(\eta)}\diff \eta^2 +  \frac{\Go(\eta)}{\eta-\xi}(\diff \Phi + \xi \diff \Psi)^2 ~,
\ee
where $\Fo(\xi)$ and $\Go(\eta)$ are a priori arbitrary functions. Note that $\partial/\partial \Phi$, $\partial/\partial\Psi$ are Killing vector fields, hence this is a co-homogeneity two metric. 
The K\"ahler form has a universal expression, independent of $\Fo(\xi)$, $\Go(\eta)$:
\be
X^1 \ =\ \frac{1}{g^2}\,\diff \left[ (\eta + \xi )\diff \Phi + \eta\xi \,\diff \Psi \right]~.
\ee
The term orthotoric means that the momentum maps $\eta + \xi$, $\eta \xi$ for
the Hamiltonian Killing vector fields $\partial/\partial \Phi$, $\partial/\partial\Psi$, respectively, have the property that the one-forms $\diff \xi$, $\diff \eta$ are orthogonal. As a consequence, the K\"ahler metric does not contain a $\diff \eta \diff \xi$ term. 

It is convenient to introduce an orthonormal frame
\bea
&&  E_1 = \frac{1}{g}\sqrt{\frac{\eta-\xi}{\Fo(\xi)}}\,\diff \xi ~,   \qquad   E_2 =  \frac{1}{g}\sqrt{\frac{\Fo(\xi)}{\eta-\xi}}\,(\diff \Phi + \eta \diff \Psi) ~, \nonumber\\
 && E_3 = \frac{1}{g}\sqrt{\frac{\eta-\xi}{\Go(\eta)}}\, \diff \eta~, \qquad E_4 =  \frac{1}{g}\sqrt{\frac{\Go(\eta)}{\eta-\xi}}\,(\diff \Phi + \xi \diff \Psi)  ~,
\eea
with volume form vol$_B= - E_1\wedge E_2 \wedge E_3 \wedge E_4$. 
 Then the K\"ahler form can be written as
\be
X^1 \ =\ E_1 \wedge E_2 + E_3 \wedge E_4~.
\ee
For the complex two-form $\Omega$ we can take
\be  \label{Omega}
\Omega\ =\ X^2 + i X^3\ =\ (E_1 - i E_2)\wedge (E_3 - i E_4 )\ .
\ee
This satisfies the properties~\eqref{diffOmega},~\eqref{Xalgebra}, with the Ricci form potential given by
\be
P \ =\ \frac {\Fo'(\xi) (\diff \Phi + \eta \diff \Psi )  + \Go'(\eta)  (\diff \Phi + \xi \diff \Psi )  }{2(\xi- \eta)}~.
\label{orthoriccipotential}
\ee
Other formulae that we will need are the Ricci scalar
\be\label{OrthoRicciScal}
R \ =\  g^2\,\frac{\Fo''(\xi)+ \Go'' (\eta)}{\xi-\eta} \ ,
\ee
and its Laplacian
\be\label{orthoLaplacianR}
\nabla^2 R\ =\ \frac{g^2}{\eta -\xi}\left[ \partial_\xi ( \Fo\, \partial_\xi R )  +\partial_\eta ( \Go\, \partial_\eta R )    \right]\ .
\ee

\subsection{The solution}

To construct the solution we plug our orthotoric ansatz in the supersymmetry equations of section~\ref{GGclassification}.
Eq.~\eqref{solf} gives for the function $f$
\be\label{orthof}
f \ =\  \frac{24(\eta-\xi)}{\Fo''(\xi) + \Go''(\eta)}\ .
\ee
In order to solve eq.~\eqref{G+plusG-} for $\omega$, we need to first construct $G^+$, $G^-$. From eq.~\eqref{solG+} we obtain
\be
G^+ \ =\ \frac{1}{8g}(\partial_\xi \Ho - \partial_\eta \Ho) \left( E_1 \wedge E_2 - E_3 \wedge E_4 \right)\ ,
\ee 
where we introduced the useful combination
\be
\Ho(\eta,\xi) \ =\ g^2\,\frac{\Fo'(\xi) + \Go'(\eta)}{\eta - \xi}\ .
\ee
We recall that $G^- = \frac{1}{2gR} \sum_{I=1}^3 \lambda^I X^I$, and we have to compute the functions $\lambda^1,\lambda^2,\lambda^3$.
Eq.~\eqref{sollambda1} gives
\be
\lambda^1 \ =\  \frac 12 \nabla^2 R -\frac{2}{3}\,\partial_\xi \Ho\,\partial_\eta \Ho\ ,
\ee
where $\nabla^2R$ was expressed in terms of orthotoric data above. In order to solve for $\lambda^2,\lambda^3$, we have to analyse the constraint~\eqref{constraintfinal} on the K\"ahler metric. Plugging our ansatz in, we obtain the equation
\bea
&&\!\!\partial_\xi \left[ \Fo\,\partial_\xi \Ho\, \partial_\xi (\partial_\xi \Ho  + \partial_\eta \Ho)  + \Fo\,\partial_\xi  \left(\nabla^2 R -\tfrac{4}{3}\,\partial_\xi \Ho\,\partial_\eta \Ho \right) \right] \qquad\nn \\ [2mm]
&+&\!\!\partial_\eta \left[ \Go\,\partial_\eta \Ho\, \partial_\eta (\partial_\xi \Ho  + \partial_\eta \Ho)  + \Go\,\partial_\eta  \left(\nabla^2 R -\tfrac{4}{3}\,\partial_\xi \Ho\,\partial_\eta \Ho \right) \right] 
 \ = \ 0\ .
 \label{orthomaster}
\eea
This is a complicated sixth-order equation for the two functions $\Fo(\xi)$ and $\Go(\eta)$, that we have not been able to solve in general. However, we have found the cubic polynomial solution
\bea\label{solFG}
\Go(\eta) &=& g_4(\eta - g_1)(\eta - g_2)(\eta - g_3) \ , \nn \\ [2mm]
\Fo(\xi) &=& -\Go(\xi) + f_1(\xi + f_0)^3\ ,
\eea
comprising six arbitrary\footnote{This includes the case where $g_4\to 0$ and one or more roots diverge, so that the cubic $\Go$ degenerates to a polynomial of lower degree. Same for $\Fo$.} parameters $g_1,\dots,g_4$, $f_0$, $f_1$.
We thus continue assuming that $\Fo$ and $\Go$ take the form~\eqref{solFG}. We can then solve eq.~\eqref{eqforlambda23real} for $\lambda^2, \lambda^3$. Assuming a dependence on $\eta, \xi$ only, the solution is
\be\label{sollambda23}
\lambda^2 + i \lambda^3 \  = \ i\,g^4\, 
 \frac{\Fo'''+\Go'''}{(\eta-\xi)^3} \sqrt{\Fo(\xi) \Go(\eta)} \,+  g^4\frac{c_2+ i c_3}{\sqrt{\Fo(\xi) \Go(\eta)}}~,
\ee
with $c_2,c_3$ real integration constants. One can promote $c_2+ic_3$ to an arbitrary anti-holomorphic function, however we will not discuss such generalisation in this paper (see~\cite{Figueras:2006xx} for an example where this has been done explicitly).

We now have all the ingredients to solve eq.~\eqref{G+plusG-} and determine $\omega$. 
The solution is
\bea\label{solomega}
\omega \!\!&=&\!\! \frac{\Fo'''+\Go'''}{48g(\eta-\xi)^2}\Big\{\left[ \Fo(\xi) + (\eta-\xi) \left(\tfrac 12 \Fo'(\xi) -\tfrac 14 \Fo'''(\xi)(f_0+\xi)^2\right) \right](\diff \Phi + \eta \diff \Psi) \nn \\ [2mm] 
\!\!&+&\!\!\! \Go(\eta) (\diff \Phi + \xi \diff \Psi)  \Big\} 
\,-\, \frac{\Fo'''\Go'''}{288g} \left[ (\eta+\xi)\diff \Phi + \eta\xi\, \diff \Psi \right]\nn \\ [2mm]
\!\!&-&\!\!\! \frac{c_2}{48g}\Big(I_1\frac{\xi\diff \xi}{\Fo(\xi)} + I_2 \frac{\eta\diff \eta}{\Go(\eta)} +\Phi \diff \Psi \Big) - \frac{c_3}{48g} \left[ (I_1-I_2) \diff \Phi + (I_3-I_4) \diff \Psi   \right] + \diff \chi \ ,\qquad
\eea
where
\be
I_1 = \int \frac{\diff\eta}{\Go(\eta)} \ , \qquad  I_2  = \int \frac{\diff\xi}{\Fo(\xi)} \ , \qquad
I_3 = \int \frac{\eta\,\diff\eta}{\Go(\eta)} \ , \qquad I_4  = \int \frac{\xi\,\diff\xi}{\Fo(\xi)} \ .
\ee
Moreover, $\diff \chi$ is an arbitrary locally exact one-form, which in the five-dimensional metric can be reabsorbed by a change of the $y$ coordinate.
For $\Fo$ and $\Go$ as in~\eqref{solFG}, the integrals $I_1,\ldots,I_4$ can be expressed in terms of the roots of the polynomials. We have:
\be
I_1 = \frac{\log(\eta - g_1)}{g_4(g_1 - g_2)(g_1-g_3)} + {\rm cycl(1,2,3)}\ , \qquad I_3 = \frac{g_1\log(\eta - g_1)}{g_4(g_1 - g_2)(g_1-g_3)} + {\rm cycl(1,2,3)}\ ,
\ee
and similarly for $I_2$ and $I_4$ (although the roots of $\Fo$ in~\eqref{solFG}  expressed in terms of the parameters  $g_1,\dots,g_4$, $f_0$, $f_1$ are less simple). Here, ${\rm cycl(1,2,3)}$ denotes cyclic permutations of the roots.

Note that if $c_2\neq 0$ then $\omega$ explicitly depends on one of the angular coordinates $\Phi,\Psi$, hence the $U(1)\times U(1)$ symmetry of the orthotoric base is broken to a single $U(1)$ in the five-dimensional metric.

Also note that when $\Fo'''+\Go''' = 0$, namely the coefficients of the cubic terms in the polynomials are opposite, expression \eqref{solomega} simplifies drastically. Then we see that the term in $\diff \omega $ independent of $c_2,c_3$ is proportional to the K\"ahler form $X^1$. Moreover the base becomes K\"ahler-Einstein.  
This class of solutions was pointed out
in~\cite{Gauntlett:2003fk}, where it was explored for the case the base is the space $SU(2,1)/S(U(2)\times U(1))$ endowed with the Bergmann metric. This is the non-compact analog of $\mathbb CP^2$ with the Fubini-Study metric, and the corresponding solution with $c_1=c_2=0$ is pure AdS$_5$.

To summarise, we started from the orthotoric ansatz~\eqref{orthometric} for the four-dimensional K\"ahler metric, studied the sixth-order constraint~\eqref{constraintfinal} and found a solution in terms of cubic polynomials $\Fo$, $\Go$ containing six arbitrary parameters, {\it cf.} \eqref{solFG}. We also provided explicit expressions for $P$, $f$ and $\omega$ ({\it cf.}\ \eqref{orthoriccipotential}, \eqref{orthof}, \eqref{solomega}), with the solution for $\omega$ containing the additional parameters $c_2,c_3$. Plugging these expressions in the metric~\eqref{5dMetricTimelike} and Maxwell field~\eqref{susygaugefield}, we thus obtain a supersymmetric solution to minimal gauged supergravity controlled by eight parameters. We now show that three of the six parameters in the polynomials are actually trivial in the five-dimensional solution.

\subsection{Triviality of three parameters}\label{TrivialitySec}

As a first thing, we observe that one is always free to rescale the four-dimensional K\"ahler base by a constant factor. This is because the spinor solving the supersymmetry equation~\eqref{KillingSpEqDirac} is defined up to a multiplicative constant, and the spinor bilinears inherit such rescaling freedom. This leads to the transformation
\bea
&& X^{I} \,\to\, \const\, X^{I}\ , \qquad f \to \const\, f \ , \qquad y \to \const^{-1} y\ ,\nn \\ [1mm]
&&\diff s^2_B \to \const\, \diff s^2_B  \ ,\qquad P \to P\ ,\qquad   \omega \to \const^{-1}\omega\ ,
\eea
where $\const$ is a non-zero constant. Clearly this leaves the five-dimensional metric~\eqref{5dMetricTimelike} and the gauge field \eqref{susygaugefield} invariant.

Let us now consider a supersymmetric solution whose K\"ahler base metric $\diff s^2_B$ is in the orthotoric form \eqref{orthometric}, with some given functions $\Fo(\xi)$ and $\Go(\eta)$. Then we can use the symmetry above to rescale these two functions. Indeed after performing the transformation we have $ (\diff s_B^2)^{\rm old} = \const \,(\diff s^2_B)^{\rm new}$, and the new K\"ahler metric is again in orthotoric form, with the redefinitions
\be\label{triviality1}
 \Fo^{\rm old} = \const^{-1} \Fo^{\rm new} \ , \qquad \Go^{\rm old} = \const^{-1} \Go^{\rm new} \ , \qquad \Phi^{\rm old} = \const\,\Phi^{\rm new} \ , \qquad \Psi^{\rm old} = \const\, \Psi^{\rm new} \ .
\ee
Hence the overall scale of $\Fo$ and $\Go$ is irrelevant as far as the five-dimensional solution is concerned.
A slightly more complicated transformation that we can perform is
\bea
&& \xi^{\rm old} = \const_2 \xi^{\rm new} + \const_3 \ , \qquad\qquad\qquad \eta^{\rm old} = \const_2 \eta^{\rm new} + \const_3\nn \ ,\\ [2mm] 
&& \Psi^{\rm old} = \const_1 \const_2 \Psi^{\rm new}\ , \;\:\qquad\qquad \qquad \Phi^{\rm old} = \const_1(\const_2^2\Phi^{\rm new} - \const_2\const_3 \Psi^{\rm new})\ ,\nn \\ [2mm]
&& \Fo^{\rm old}(\xi^{\rm old}) =  \const_1^{-1} \Fo^{\rm new}(\xi^{\rm new})\ , \qquad \Go^{\rm old}(\eta^{\rm old}) = \const_1^{-1} \Go^{\rm new}(\eta^{\rm new})\ .
\eea
with arbitrary constants $\const_1\neq 0$, $\const_2 \neq 0$ and $\const_3$, such that $\const_1 \const_2^3 = \const$.
It is easy to see that the new metric $(\diff s^2_B)^{\rm new}$ is again orthotoric, though with different cubic functions $\Fo$ and $\Go$ compared to the old ones.

We conclude that a supersymmetric solution with orthotoric K\"ahler base is locally equivalent to another orthotoric solution, with functions 
\be
\Fo^{\rm new }(\xi) = \const_1 \Fo^{\rm old}(\const_2\xi +\const_3)\ , \qquad \Go^{\rm new}(\eta)= \const_1 \Go^{\rm old}(\const_2\eta + \const_3)\ .
\ee
Using this freedom, we can argue that three of the six parameters in our orthotoric solution are trivial. 
 In the next section we will show that the remaining ones are not trivial by relating our solution with $c_2=c_3=0$ to  the solution of~\cite{Chong:2005hr}.

\subsection{Relation to the solution of~\cite{Chong:2005hr}}\label{sec:relationCCLP}

The authors of \cite{Chong:2005hr} provide a four-parameter family of AAdS solutions to minimal five-dimensional gauged supergravity. 
 The generic solution preserves $U(1)\times U(1)\times \mathbb R$ symmetry (where $\mathbb R$ is the time direction) and is non-supersymmetric. By fixing one of the parameters, one obtains a family of supersymmetric solutions, controlled by the three remaining parameters $a,b,m$. This includes the most general supersymmetric black hole free of closed timelike curves (CTC's) known in minimal gauged supergravity, as well as a family of topological solitons. 
Generically, the supersymmetric solutions are 1/4 BPS in the five-dimensional theory, namely they preserve two real supercharges.
For $b=a$ or $b=-a$, the symmetry is enhanced to $SU(2)\times U(1)\times \mathbb R$.

We find that upon a change of coordinates the supersymmetric solution of \cite{Chong:2005hr} fits in our orthotoric solution, with polynomial functions $\Fo$, $\Go$ of the type discussed above. In detail, the five-dimensional metric and gauge field strength of~\cite{Chong:2005hr} match~\eqref{5dMetricTimelike},~\eqref{susygaugefield}, with the data given in the previous section and $c_2=c_3=0$. The change of coordinates is
\bea\label{CCLPtoOrtho1}
t_{\rm CCLP} & = & y\nn\\ [1mm]
\theta_{\rm CCLP} &=& \frac 12\arccos\eta \nn \\ [1mm]
r^2_{\rm CCLP} &=& \frac{1}{2}(a^2-b^2)\tilde m\, \xi + \frac{1}{g}\left[(a+b)\tilde m + a+b+abg \right] + \frac 12 (a+b)^2\tilde m \ ,\quad \nn \\ [1mm]
\phi_{\rm CCLP} &=& g\,y -4\frac{1-a^2g^2}{(a^2-b^2)g^2\tilde m}\,(\Phi-\Psi) \ ,\nn \\ [1mm] 
\psi_{\rm CCLP} &=& g\,y -4\frac{1-b^2g^2}{(a^2-b^2)g^2\tilde m}\,(\Phi+\Psi)\ ,
\eea
where ``CCLP'' labels the coordinates of~\cite{Chong:2005hr}.
Here, we found convenient to trade $m$ for
\be\label{defmtilde}
\tilde m \ =\ \frac{m \,g}{(a+b)(1+ag)(1+bg)(1+ag+bg)} -1\,,
\ee
which is defined so that the black hole solution of~\cite{Chong:2005hr} corresponds to $\tilde m = 0$.
The cubic polynomials $\Fo(\xi)$ and $\Go(\eta)$ read
\bea\label{CCLPtoOrtho2}
\Go(\eta)&=&-\frac{4}{(a^2-b^2)g^2\tilde m}(1-\eta^2) \left[ (1-a^2g^2)(1+\eta) + (1-b^2g^2)(1-\eta) \right] \ ,\nn \\ [2mm]
\Fo(\xi) &=& - \Go(\xi) - 4\,  \frac{1+\tilde m}{\tilde m}\left( \frac{2+ag+bg}{(a-b)g} + \xi \right)^3\ ,
\eea
and are clearly of the form~\eqref{solFG}.\footnote{Note that the present orthotoric form of the solution in \cite{Chong:2005hr}, which is adapted to supersymmetry, does not use the same coordinates of the Pleba\'nski-Demia\'nski-like form appearing in~\cite{Kubiznak:2009qi}.} The function $\chi$ in~\eqref{solomega} is $\chi = -\frac{2\Psi}{g\tilde m}$.
The Killing vector arising as a bilinear of the spinor $\epsilon$ solving the supersymmetry equation~\eqref{KillingSpEqDirac} is
\be\label{susyV_CCLP}
V \ = \ \frac{\partial}{\partial y} \ =\ \frac{\partial}{\partial t_{\rm CCLP}} + g \frac{\partial}{\partial \phi_{\rm CCLP}} + g \frac{\partial}{\partial \psi_{\rm CCLP}}\ .
\ee

Combining the arguments of section~\ref{TrivialitySec} with the map just presented, we conclude that for $c_2=c_3=0$, the family of supersymmetric solutions we have constructed is (at least locally) equivalent to the supersymmetric solutions of~\cite{Chong:2005hr}. 

We checked that when either $c_2$ or $c_3$ (or both) are switched on, the boundary metric is no more conformally flat, hence the solution becomes AlAdS and is not diffeomorphic to the $c_2 = c_3 =0$ case. We have thus obtained a \emph{new} two-parameter AlAdS deformation of the AAdS solutions of~\cite{Chong:2005hr}. Choosing $c_2\neq 0, c_3 = 0$ and $\chi$ in~\eqref{solomega} as $\chi=-\frac{2\Psi}{g\tilde m}+ \frac{c_2}{48g}I_1I_4$, the boundary metric appears to be regular and of type Petrov III like that of~\cite{Gauntlett:2003fk,Gauntlett:2004cm}.\footnote{See~\cite{Cassani:2013dba} for a discussion of the Petrov type of supersymmetric boundaries.} Its explicit expression in the coordinates of~\cite{Chong:2005hr} is (below we drop the label ``CCLP'' on the coordinates):
\be
\diff s^2_{\rm bdry} \ = \ \diff s^2_{\rm bdry,CCLP} + \diff s^2_{c_2}\ ,
\ee
where the undeformed boundary metric of~\cite{Chong:2005hr}, obtained sending $gr\to \infty$, is
\be\label{bdrymetrCCLP}
\diff s^2_{\rm bdry,CCLP} \ =\ -\frac{ \Delta_\theta}{\Xi_a \Xi_b}\dd t^2+\frac{1}{g^2} \left( \frac{\dd \theta^2}{\Delta_\theta }+\frac{ \sin^2 \theta}{\Xi_a} \dd \phi^2 +\frac{ \cos ^2\theta}{\Xi_b}\dd \psi^2  \right),
\ee
with $\Xi_a = 1-a^2g^2$, $\Xi_b = 1-b^2g^2$ and 
\be\label{Deltatheta}
\Delta_{\theta}=1-a^2g^2\cos^2\theta-b^2g^2\sin^2\theta\ ,
\ee
while the deformation is linear in $c_2$ and reads
\bea
\diff s^2_{c_2}
 \!\!\!&=&\!\!\! c_2\frac{ g^2 \tilde m^2 \left(a^2-b^2\right)^2 }{1536  \Xi_a^3 \Xi_b^3 }  \big(gt(\Xi_a+\Xi_b )- \Xi_b \phi -\Xi_a \psi   \big) \big(-g \dd t  \left(\Xi_a -\Xi_b \right) -  \Xi_b \dd \phi+\Xi_a \dd \psi \big)\nn\\
\!\!\!&&\quad \times  \left(-  (  \Xi_a \cos^2 \theta +\Xi_b \sin^2 \theta ) g \dd t +\Xi_a \cos^2 \theta \dd \psi+\Xi_b \sin^2 \theta \dd \phi     \right) \ .
\eea
It would be interesting to study further the regularity properties of these deformations and see if they generalise the similar solutions of~\cite{Gauntlett:2003fk,Gauntlett:2004cm,Figueras:2006xx}.

Note that both the change of coordinates~\eqref{CCLPtoOrtho1} and the polynomials~\eqref{CCLPtoOrtho2} are singular in the limits $\tilde m \to 0$ or $b\to a$, while they remain finite when $b\to -a$. (When we take $b\to \pm a$, it is understood that we keep $m$, and not $\tilde m$, fixed).
As already mentioned, these are physically relevant limits: $\tilde m \to 0$ defines the black hole solutions free of CTC's, while $b\to \pm a$ leads to solutions with enhanced symmetry. We clarify the singular limits in the next section.

\section{A scaling limit and two special cases}\label{sec:scalinglimit}

In the following we show that a simple scaling limit of the orthotoric metric yields certain {\it non}-orthotoric K\"ahler metrics, that have previously been employed to construct supersymmetric solutions. We recover on the one hand the base metric considered in~\cite{Figueras:2006xx}, and on the other hand an $SU(2)\times U(1)$ invariant K\"ahler metric. This proves that our orthotoric ansatz captures all known supersymmetric solutions to minimal five-dimensional gauged supergravity belonging to the timelike class.
The procedure will also clarify the singular limits pointed out in the previous section. 

We start by redefining three of the four orthotoric coordinates $\{\eta,\xi,\Phi,\Psi\}$ as
\be
\Phi = \varepsilon\, \phi\ , \qquad \Psi = \varepsilon\,\psi\ , \qquad \xi = - \varepsilon^{-1}\,\rho\ ,
\ee
where $\varepsilon$ is a parameter that we will send to zero. For the metric to be well-behaved in the limit, we also assume that the functions $\Fo$, $\Go$ satisfy
\be
\Go(\eta) = \varepsilon^{-1}\widetilde{\Go}(\eta) + \mathcal O(1)\ , \qquad \Fo(\xi)  = \varepsilon^{-3} \widetilde{\Fo}(\rho) + \mathcal O(\varepsilon^{-2})\ ,
\ee
where $\widetilde{\Go}(\eta)$, $\widetilde{\Fo}(\rho)$ are independent of $\varepsilon$ and thus remain finite in the limit.
Plugging these in the orthotoric metric \eqref{orthometric} and sending $\varepsilon \to 0$ we obtain
\be\label{limitorthometric}
g^2\diff s^2_B \ = \ g^2\lim_{\varepsilon \to 0} \,\diff s^2_{\rm ortho} \ = \ \frac{\rho}{\widetilde{\Fo}(\rho)}\diff\rho^2 + \frac{\widetilde{\Fo}(\rho)}{\rho}(\diff \phi + \eta\, \diff \psi)^2 + \rho\bigg(\frac{\diff \eta^2}{\widetilde \Go(\eta)} + \widetilde \Go(\eta) \diff\psi^2\bigg)\ .
\ee
This is a K\"ahler metric of Calabi type (see \emph{e.g.} \cite{ACGweakly}), with associated K\"ahler form
\be
X^1 \ = \ -\,\frac{1}{g^2}\,\diff\left[ \rho (\diff \phi +\eta \diff \psi) \right]\ .
\ee
At this stage the functions $\widetilde{\Fo}(\rho)$ and $\widetilde \Go(\eta)$ are arbitrary. Of course, for~\eqref{limitorthometric} to be the base of a supersymmetric solution we still need to impose   on   $\widetilde{\Fo}(\rho)$,  $\widetilde \Go(\eta)$ the equation following from the constraint~\eqref{constraintfinal}. 

We next consider two subcases: in the former we fix $\widetilde{\Fo}$ and recover the metric studied in~\cite{Figueras:2006xx}, while in the latter we fix $\widetilde{\Go}$ and obtain an $SU(2)\times U(1)$ invariant metric.

\paragraph{Case 1.} We take $\widetilde{\Fo}(\rho) = 4\rho^3 + \rho^2$ and subsequently redefine $\rho = \frac{1}{4}\sinh^2(g\sigma)$. Then \eqref{limitorthometric} becomes
\be
\diff s^2_B \ = \ \diff \sigma^2 + \frac{1}{4g^2}\sinh^2(g\sigma)\bigg(  \frac{\diff \eta^2}{\widetilde \Go(\eta)} + \widetilde \Go(\eta) \diff\psi^2 + \cosh^2(g\sigma) (\diff\phi + \eta \diff\psi)^2 \bigg)\ ,
\ee
which is precisely the metric appearing in eq.~(7.8) of~\cite{Figueras:2006xx} (upon identifying $\eta = x$ and $\widetilde \Go(\eta) = H(x)$). In this case our equation~\eqref{constraintfinal} becomes
\be\label{integrabilityTildeG}
\big(\widetilde{\Go}^2\, \widetilde{\Go}''''\big)'' \, = \ 0\ ,
\ee
that coincides with the constraint found in~\cite{Figueras:2006xx}.
As discussed in~\cite{Figueras:2006xx}, this K\"ahler base metric supports the most general  supersymmetric black hole solution free of CTC's that is known within minimal five-dimensional gauged supergravity. This is obtained from the supersymmetric solutions of~\cite{Chong:2005hr} by setting $\tilde m =0$. In fact, the limit $\tilde m \to 0$ in the map~\eqref{CCLPtoOrtho1}, \eqref{CCLPtoOrtho2} is an example of the present $\varepsilon \to 0$ limit, where the resulting $\widetilde \Go(\eta)$ is a cubic polynomial~\cite{Kunduri:2006ek,Figueras:2006xx}.\footnote{This can be seen starting from~\eqref{CCLPtoOrtho1}, \eqref{CCLPtoOrtho2} and redefining $\tilde m=-\frac{8\alpha^2}{(a^2-b^2)}\varepsilon$ and $r^2= r_0^2+ 4\alpha^2 \rho$, where we are denoting $\alpha^2 = r_0^2+\frac{(1+ag+bg)^2}{g^2}$ and $r_0^2 = \frac{a+b+abg}{g}$. It follows that $\xi = \varepsilon^{-1}\rho + \mathcal O(1)$. Then implementing the scaling limit described above we get $\widetilde{\Fo}(\rho) = 4\rho^3 + \rho^2$ and $\widetilde{\Go}(\eta) = \frac{1}{2}(1-\eta^2)\left[ A_1^2 + A_2^2 + (A_1^2-A_2^2) \eta \right]$ with $A_1^2 = \frac{1-a^2g^2}{g^2\alpha^2}$ and $A_2^2 = \frac{1-b^2g^2}{g^2\alpha^2}$. This makes contact with the description of the supersymmetric black holes of~\cite{Chong:2005hr} given in~\cite{Kunduri:2006ek,Figueras:2006xx}.} Particular non-polynomial solutions to eq.~\eqref{integrabilityTildeG} were found in~\cite{Figueras:2006xx}, however in the same paper these were shown to yield unacceptable singularities in the five-dimensional metric.

\paragraph{Case 2.} If instead we take $\widetilde{\Go}(\eta) = 1-\eta^2$ and redefine $\eta = \cos\theta$, then the metric \eqref{limitorthometric} becomes
\be\label{SU2limitorthometric}
g^2\,\diff s^2_B \ = \ \frac{\rho}{\widetilde{\Fo}(\rho)}\diff\rho^2 + \frac{\widetilde{\Fo}(\rho)}{\rho}(\diff \phi + \cos\theta\, \diff \psi)^2 + \rho\left(\diff \theta^2 + \sin^2\theta \diff\psi^2\right)\ ,
\ee
with K\"ahler form
\be
X^1 \ = \ - \,\frac{1}{g^2}\,\diff \left[\rho (\diff \phi + \cos\theta \diff \psi) \right]\ .
\ee
This has enhanced $SU(2)\times U(1)$ symmetry compared to the $U(1)\times U(1)$ invariant orthotoric metric. 
It is in fact the most general K\"ahler metric with such symmetry and is equivalent, by a simple change of variable, to the metric ansatz employed in~\cite{Gutowski:2004ez} to construct the first supersymmetric AAdS black hole free of CTC's. The constraint~\eqref{constraintfinal} becomes a sixth-order equation for $\widetilde{\Fo}(\rho)$. This is explicitly solved if $\widetilde{\Fo}(\rho)$ satisfies the fifth-order equation
\bea
&&16 (\ti{\Fo}')^2 + 4\rho^2 \left(6\ti{\Fo}''+ (\ti{\Fo}'')^2 -2\rho\ti{\Fo}^{(3)}\right) + 2\rho\ti {\Fo}'\left(-24-4\ti {\Fo}'' - 4\rho\ti {\Fo}^{(3)} +3\rho^2 \ti {\Fo}^{(4)}\right) \nn \\ [2mm]
&&-3\ti {\Fo} \left( -16 + 8 \ti {\Fo}'' - 8\rho \ti {\Fo}^{(3)} + 4\rho^2 \ti {\Fo}^{(4)} - \rho^3\ti {\Fo}^{(5)} \right) \ = \ 0\ .\label{SU2U1invariantEq}
\eea
Upon a change of variable, the latter is equivalent to the sixth-order equation presented in~\cite[eq.$\:$(4.23)]{Gutowski:2004ez}. It was proved there that a solution completely specifies an $SU(2)\times U(1)$ invariant five-dimensional metric and graviphoton.
We find that a simple solution to~\eqref{SU2U1invariantEq} is provided by a cubic polynomial 
\be\label{PolySolTildeF}
\widetilde{\Fo}(\rho) = f_0 + f_1 \rho + f_2 \rho^2 + f_3 \rho^3\ , \quad{\rm such\ that}\quad  f_1^2 + 3 f_0 (1 - f_2)=0\ .
\ee 
Supersymmetric AAdS solutions with $SU(2)\times U(1)$ symmetry were also found in~\cite{Cvetic:2004hs} and further discussed in~\cite{Cvetic:2005zi}. It is easy to check that after scaling away a trivial parameter, the five-dimensional solution determined by~\eqref{PolySolTildeF} in fact reproduces\footnote{In the case the charges are set equal, so that the two vector multiplets of the $U(1)^3$ gauged theory can be truncated away and the solution exists within minimal gauged supergravity.} the two-parameter ``case B'' solution given in \cite[sect.$\:$3.4]{Cvetic:2005zi}. In turn, the latter includes the black hole of~\cite{Gutowski:2004ez}, and a family of topological solitons for particular values of the parameters. 

The special case $f_1=0$, $f_2=1$ yields the most general K\"ahler-Einstein metric with $SU(2)\times U(1)$ isometry; this has curvature $R = -6g^2f_3$ and is diffeomorphic to the Bergmann metric only for $f_0=0$. The corresponding $SU(2)\times U(1)$ invariant five-dimensional solution is ``Lorentzian Sasaki-Einstein'': for $f_0=0$ this is just AdS$_5$, while for $f_0\neq0$ it features a curvature singularity at $\rho=0$.

In~\cite{Cassani:2014zwa}, a different solution of equation~\eqref{SU2U1invariantEq} was put forward, leading to a smooth AlAdS five-dimensional metric. The non-conformally flat boundary is given by a squashed $S^3\times \mathbb R$, where the squashing is along the Hopf fibre and thus preserves $SU(2)\times U(1)$ symmetry.

A particular example of this $\varepsilon \to 0$ limit is given by the $b\to a$ limit in the map~\eqref{CCLPtoOrtho1}, \eqref{CCLPtoOrtho2} relating the solution of~\cite{Chong:2005hr} and the one based on our orthotoric ansatz.\footnote{This can be seen starting from~\eqref{CCLPtoOrtho1}, \eqref{CCLPtoOrtho2}, redefining $b= a + 8(1-a^2g^2)\big[\frac{g^3m}{(1+2ag)(1+ag)^2}-2ag^2\big]^{-1}\varepsilon$ after having re-expressed $\tilde m$ as in \eqref{defmtilde}, and implementing the scaling limit. This gives $\widetilde{\Go}(\eta) = 1-\eta^2$ and a cubic polynomial $\widetilde{\Fo}(\rho)$ satisfying \eqref{PolySolTildeF}.}
In fact, taking $b=a$ in the solutions of~\cite{Chong:2005hr} yields precisely the solutions presented in~\cite[sect.$\:$3.4]{Cvetic:2005zi}.

Note that since the black hole of~\cite{Gutowski:2004ez} is obtained from the general solution of~\cite{Chong:2005hr} by taking $\tilde m =0$ and $b=a$, it belongs both to our cases 1 and 2.

\begin{figure}
	\centering
\includegraphics[width=15cm]{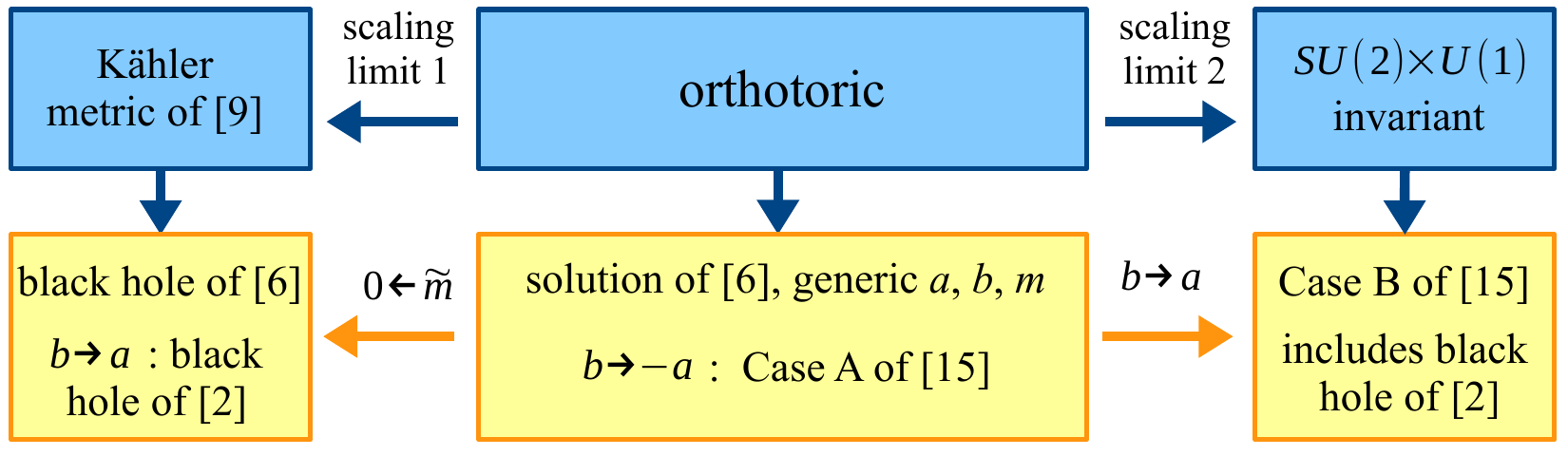}
\caption{K\"ahler base metrics (above) and corresponding known AAdS solutions (below), with relevant references.}
\label{Diagram}
\end{figure}

In figure~\ref{Diagram} we summarise the relation between different K\"ahler metrics and the corresponding AAdS solutions in five dimensions.

\section{Topological solitons}\label{sec:toposoli}

In this section we focus on a sub-family of the solution of~\cite{Chong:2005hr}, comprising  ``topological solitons'' with non-trivial geometry but no horizon. \emph{A priori}, these may 
 be considered as candidate gravity dual to pure states of SCFT's defined on $S^3\times \mathbb R$. In section~\ref{sec:compareCasimir} we consider the non-vanishing vacuum expectation values of the energy and $R$-charge of such theories, and we look for a possible gravity dual. The constraints from the superalgebra naturally lead us to consider a 1/2 BPS topological soliton, however a direct comparison of the charges with the SCFT vacuum expectation values shows that these do not match. In section~\ref{12remarks} we argue that 
 in the dual SCFT certain background $R$-symmetry field must be turned on,  implying a constraint on the $R$-charges and suggesting that the state dual to the topological soliton is different from the vacuum.
 Finally, in section~\ref{sec:1/4BPStopsol} we show that the one-parameter family of  1/4 BPS topological solitons presented in~\cite{Chong:2005hr} is  generically plagued with conical singularities.

\subsection{Comparison with the supersymmetric Casimir energy}\label{sec:compareCasimir}

In this section we assess the possible relevance of the supergravity solutions discussed above to account for the vacuum state of dual four-dimensional $\mathcal N=1$ SCFT's defined on the cylinder $S^3 \times \mathbb R$ \cite{Kim:2012ava,Assel:2014paa,Assel:2015nca,Lorenzen:2014pna,Bobev:2015kza}. 
The field theory background is specified by a round metric on $S^3$ with radius $r_3$, and by a flat connection for a non-dynamical gauge field $A^{\rm cs}$ coupling to the $R$-current.\footnote{The label ``cs'' refers to the fact that this is the gauge field of the four-dimensional conformal supergravity that determines how the SCFT is coupled to curved space.} Crucially, $A^{\rm cs}$ is chosen in such a way that half of the eight supercharges in the superconformal algebra commute with the Hamiltonian generating time translations on the cylinder. In this way we ensure that this half of the supercharges is preserved when the Euclidean time on the cylinder is compactified to a circle. The Hamiltonian, that we will denote by $H_{\rm susy}$, is related to the operator $\Delta$ generating dilatations in flat space as $H_{\rm susy} = \Delta - \frac{1}{2r_3}R$, where $R$ is the $R$-charge operator (see \cite{Assel:2015nca} for more details). 
We will call $\mathcal Q_\alpha$, $\mathcal Q^\dagger{}^\alpha$, $\alpha=1,2$ the conserved supercharges. They transform in the $({\bf 2}, {\bf 1})$ representation of the $SU(2)_{\rm left}\times SU(2)_{\rm right}$ group acting on $S^3$, and their anti-commutator is~\cite{Romelsberger:2005eg} 
\be
\label{supali} \frac 12 \{\mathcal Q_\alpha,  \mathcal Q^{\dagger}{}^{\beta}\}  \ =\  \delta{}^{\beta}{}_{\alpha}  \Big(  H_{\rm{susy}}-\frac{1}{r_3} R   \Big) - \frac{2}{r_3} \sigma^i \, {}^{\beta}{}_{\alpha}  J_{\rm left}^i~,
\ee
where the $J^i_{\rm left}$, $i=1,2,3$, generate the $SU(2)_{\rm left}$ angular momentum and $\sigma^i$ are the Pauli matrices.
The input from~\cite{Assel:2015nca} is that the vacuum  preserves all four $\mathcal Q$ supercharges, and that the bosonic charges evaluate to
\bea\label{expectedHR}
\langle H_{\rm susy} \rangle \!&\equiv&\!  \langle\Delta\rangle - \frac{1}{2r_3}\langle R\rangle  \ = \ \frac{1}{r_3}\langle R \rangle \ =\ \frac{4}{27r_3}({\bf a}+3{\bf c})\ , \nn \\ [2mm] 
\langle J_{\rm left}\rangle \!&=&\! 0 \ ,
\eea
where ${\bf a}$, ${\bf c}$ are the SCFT central charges. In~\cite{Assel:2014tba,Assel:2015nca}, these a priori divergent quantities were proved free of ambiguities as long as their regularisation does not break supersymmetry. For this proof to hold, it is important that the supercharges are preserved when the Euclidean time is compactified.

Guided by the information above, we infer that the dual supergravity solution should be AAdS and preserve (at least) four supercharges. Moreover, it should allow for a graviphoton behaving as $A \to c\, \diff t $ at the boundary, where $c$ is a constant chosen so that the asymptotic Killing spinors generating \eqref{supali} do not depend on time; in our conventions, this must be $c =-\frac{1}{\sqrt 3}$. Indeed, the general Killing spinor of AdS$_5$ that solves \eqref{KillingSpEqDirac} asymptotically reduces to a Weyl spinor on the boundary which, in standard two-component notation, may be written as
\be
\epsilon \quad \xrightarrow[r \rightarrow \infty]\quad\quad \epsilon_\alpha \ = \ (gr)^{1/2}\left( e^{\frac{i}{2}(\sqrt3 c+1)gt}\zeta_\alpha + e^{\frac{i}{2}(\sqrt3 c-1)gt}(\sigma_0\bar\eta)_\alpha \right)\ ,
\ee
where $\zeta_\alpha$ and $\bar\eta^{\dot\alpha}$ are arbitrary Weyl spinors on the $S^3 \times \mathbb R$ boundary, independent of $t$ and transforming as $({\bf 2},{\bf 1})$ and $({\bf 1},{\bf 2})$, respectively, under the action of $SU(2)_{\rm left}\times SU(2)_{\rm right}$. We see that choosing $c = -\frac{1}{\sqrt 3}$, half of the spinors become independent of time. These are the spinors that should be preserved by the solution: asymptotically, they generate the superalgebra \eqref{supali}. Note that if we Wick rotate $t\to -i\tau $ and compactify the Euclidean time $\tau$, then the other half of the Killing spinors is not well-defined. Hence we should regard Euclidean AAdS spaces (including pure AdS) with compact $S^3\times S^1$ boundary as preserving at most four supercharges.

Assuming to work in the context of type IIB supergravity on Sasaki-Einstein five-manifolds, we can translate in gravity units the value of the vacuum energy and $R$-charge given in~\eqref{expectedHR} using the standard dictionary ${\bf a}={\bf c} = \frac{\pi^2}{g^3\kappa_5^2}$. We shall also fix the radius of the boundary $S^3$ to $r_3 = 1/g$ for simplicity. Finally, we map the field theory vevs into supergravity charges as $\langle \Delta \rangle = E$, $\langle R \rangle = -\frac{1}{\sqrt 3g}Q$ and $\langle J_{\rm left}\rangle = J_{\rm left}$, where $E$ is the total gravitational energy, $Q$ the electric charge under the graviphoton and $J_{\rm left}$ the left angular momentum.
We thus obtain the following expected values for the charges of the dual gravity solution:\footnote{\label{footy} Recall that $E$ is \emph{not} the same as  $E_\mathrm{susy}=\langle H_\mathrm{susy} \rangle$, but the two quantities are related as $E =\langle \Delta \rangle = \tfrac{3}{2} \langle H_\mathrm{susy} \rangle = \tfrac{3}{2} E_\mathrm{susy}$. We hope this notation will not cause confusion in the reader.}
\be\label{ExpectedCharges}
E \ =\  -\frac{\sqrt 3}{2} Q \ =\ \frac{8}{9}\frac{\pi^2}{g^2\kappa_5^2}\ ,\qquad\qquad J_{\rm left} \ =\ 0\ .
\ee
Note that the relation between $E$ and $Q$ and the vanishing of $J_{\rm left}$ also follow  from the fact that the subset of the AdS supercharges respected by the solution anticommute into~\cite{Cvetic:2005zi}
\be\label{SugraAlgebra}
\{ \mathcal Q_{\rm sugra},{\mathcal Q}^\dagger_{\rm sugra} \} \ = \ E + \frac{\sqrt 3}{2} Q + 2g \sigma^i J^i_{\rm left}\ .
\ee
($J_{\rm right}$ instead appears in the anti-commutator of broken supercharges.)
This clearly reflects the field theory considerations above.
While there exist different prescriptions for the computation of the energy in asymptotically AdS spacetimes, here we will base our statements on the fact that this must be related to the charge $Q$ as dictated by the superalgebra. Regarding the evaluation of $Q$, we will rely on the standard formula
\be\label{formulaQcharge}
Q \ = \ \frac{1}{\kappa_5^2}\int_{S^3} *_5 F \ ,
\ee
where the integration is performed over the three-sphere at the boundary. In general this formula would contain an additional $A\wedge F$ term, however our boundary conditions impose $F\to 0$ asymptotically, hence this does not contribute to the integral.

The obvious candidate to describe the vacuum state of the dual SCFT is global AdS$_5$; indeed the boundary is $S^3 \times \mathbb R$ and we are free to switch on a graviphoton component $A_t = c$ without introducing any pathology. 
However since $F = 0$ everywhere, the charge $Q$ computed as in \eqref{formulaQcharge} obviously vanishes. A possible solution to this mismatch with~\eqref{ExpectedCharges} may come from a careful analysis of the compatibility between supersymmetry and the way the charges are evaluated. We will not address this issue here. Another option is that there may exist a solution different from empty AdS that is energetically favoured when the background gauge field is switched on. In the following we explore this possibility within the family of solutions discussed above in the paper. As we will see, the response is going to be negative, however the analysis will give us the opportunity to clarify various aspects of such solutions.
 
The requirement that the boundary be conformally flat sets $c_2=c_3=0$, hence we are left precisely with the supersymmetric solutions of~\cite{Chong:2005hr}, controlled by the three parameters $a,b,m$. In addition, the need to preserve four supercharges imposes $a+b=0$. This identifies a two-parameter family of solutions with $SU(2)\times U(1)$ invariance, originally found in~\cite{Klemm:2000vn} and further studied in \cite{Cvetic:2005zi}.\footnote{See the ``case A'' solutions in section~3.3 of \cite{Cvetic:2005zi}, with all charges set equal, $q_1=q_2=q_3=q$, so that the solution fits in minimal gauged supergravity. Refs.~\cite{Chong:2005hr,Cvetic:2005zi} base their statements on the amount of supersymmetry of the solutions on a study of the eigenvalues of the Bogomolnyi matrix arising from the AdS superalgebra. We have done a check based on the integrability condition (given in~\eqref{integrabilitycond} below) of the Killing spinor equation, and found agreement.}
As already observed, the limit $b\to -a$ (at fixed $m$) is smooth in the transformation \eqref{CCLPtoOrtho1}, \eqref{CCLPtoOrtho2} between the coordinates of~\cite{Chong:2005hr} and our orthotoric coordinates, and yields
\bea\label{mapCGLP-A_first}
t_{\rm CCLP} & = & y\ , \qquad\qquad \theta_{\rm CCLP} \ =\  \frac 12\arccos\eta \nn \\ [1mm]
r^2_{\rm CCLP} &=& \left(1-\frac{\alpha^2g^2}{q^2}\right)(\alpha g \xi +q)  -\frac{\alpha^2}{q^2} \ , \nn \\ [1mm]
\phi_{\rm CCLP} &=& g\,y -\frac{2}{\alpha g^3}\,(\Phi-\Psi)  \ , \qquad 
\psi_{\rm CCLP} \ =\ g\,y -\frac{2}{\alpha g^3}\,(\Phi+\Psi) \ ,
\eea
where we renamed the surviving parameters as
\be\label{matchCGLPAparam}
a \ =\ \frac{\alpha}{q}\ , \qquad m \ = \  \frac{(q^2 - \alpha^2g^2)^2}{q^3}  \ .
\ee
The polynomials $\Fo(\xi)$ and $\Go(\eta)$ in the orthotoric metric now become
\bea
\Go(\eta)&=&-\frac{4}{\alpha g^3}(1-\eta^2) \ ,\nn \\
\Fo(\xi) &=& - \Go(\xi) - 4 \left( \frac{q}{\alpha g} + \xi \right)^3\ .
\eea

The coordinates $\{ t,\theta,\phi,\psi,r\}$ used in~\cite{Cvetic:2005zi} are reached by the further transformation
\be\label{mapCGLP-A_second}
y  = t\ ,\ \quad 
\eta = \cos \theta\ ,\ \quad 
\xi = \frac{r^2}{\alpha g}  \ ,\ \quad
\Phi = \frac{\alpha g^3}{4}(\phi + 2gt) \ ,\ \quad
\Psi = \frac{\alpha g^3}{4}\psi \ .
\ee
In these coordinates, the five-dimensional metric reads
\be
\diff s^2_5 \ = \ - \frac{r^2 \mathcal V}{4B} \dd t^2 + \frac{\dd r^2}{\mathcal V} + B (\dd \psi + \cos \theta \dd \phi + \mathfrak f\, \dd t)^2 + \frac{1}{4} (r^2 +q) (\diff \theta^2 + \sin^2\theta\diff \phi^2) ~ ,  \label{CGLP metricA}
\ee
where
\be
\mathcal V \ = \ \frac{r^4 + g^2 (r^2+q)^3 - g^2 \alpha^2}{r^2 (r^2+q)} ~ , \qquad B \ = \ \frac{(r^2+q)^3 -\alpha^2}{4 (r^2+q)^2} \ , \qquad
\mathfrak f \ = \  \frac{2 \alpha r^2 }{\alpha^2  -(r^2+q)^3} ~,
\ee
and the graviphoton is 
\be\label{gaugefieldCGLP-A}
A \ = \  \frac{\sqrt 3}{r^2 +q} \left(  q\, \dd t - \frac{1}{2} \alpha \, (\dd \psi + \cos \theta \dd \phi) \right) +c \,\diff t\ .
\ee
Here, $\theta\in [0,\pi]$, $\phi \in [0,2\pi)$, $\psi\in [0,4\pi)$ are the standard Euler angles parameterising the three-sphere in the $S^3 \times \mathbb R$ boundary at $r = \infty$.

We observe that although the metric \eqref{CGLP metricA} and the gauge field \eqref{gaugefieldCGLP-A} are invariant under the $SU(2)_{\rm left} \times U(1)_{\rm right}$ subgroup of the $SU(2)_{\rm left} \times SU(2)_{\rm right}$ acting on the $S^3$ at infinity, the solution does not fall in the ansatz of~\cite{Gutowski:2004ez}, and hence in  case 2 of section~\ref{sec:scalinglimit}, because the bilinears of the Killing spinors and the K\"ahler base metric \eqref{orthometric} do not share the same symmetry. In particular, in the coordinates of~\cite{Cvetic:2005zi} the supersymmetric Killing vector~\eqref{susyV_CCLP} reads
\be
V \ = \ \frac{\partial}{\partial y} \ = \ \frac{\partial}{\partial t} -2g \frac{\partial}{\partial \phi}\ ,
\ee
which is invariant under $SU(2)_{\rm right}$ and transforms under $SU(2)_{\rm left}$, while the metric is invariant under $SU(2)_{\rm left} \times U(1)_{\rm right}$. In fact, the bilinears of the two independent Killing spinors of this 1/2 BPS solution give rise to three Killing vectors, which generate $SU(2)_{\rm left}$ and are $SU(2)_{\rm right}$ invariant.\footnote{It follows that the Killing vector $\frac{\partial}{\partial t} + 2g\frac{\partial}{\partial\psi}$ put forward in~\cite{Cvetic:2005zi} does not arise as a bilinear of the Killing spinors of the solution.}

We can now discuss the charges, computed using the method of~\cite{Cvetic:2005zi}. 
From~\eqref{formulaQcharge}, the charge under the graviphoton is found to be
\be\label{Qcharge}
Q \ = \ -4\sqrt 3\,q\,\frac{\pi^2}{\kappa_5^2}\ .
\ee
 The angular momentum conjugate to a rotational Killing vector $K^\mu$ is given by the Komar integral
$J = \frac{1}{2\kappa_5^2} \int_{S^3} *_5 \,\diff K \,,$ 
where $K = K_\mu \diff x^\mu$. For the  angular momentum $J_{\rm left}$ conjugate to $\frac{\partial}{\partial\phi}$ we get
\be
J_{\rm left} \ = \ 0\ ,
\ee
while $J_{\rm right}$, conjugate to $\frac{\partial}{\partial\psi}$, is controlled by $\alpha$ and reads
\be
J_{\rm right} \ = \ 2\alpha \,\frac{\pi^2}{\kappa_5^2} \ .
\ee
Finally, the energy was computed in~\cite{Cvetic:2005zi} by integrating the first law of thermodynamics, with the result
\be
E \ =\ -\frac{\sqrt 3}{2} Q \ =\  6\,q\,\frac{\pi^2}{\kappa_5^2} \ .
\ee
These values of the charges are in agreement with the superalgebra~\eqref{SugraAlgebra}.
It thus remains to check the numerical value of $Q$ against the expected one in \eqref{ExpectedCharges}. Whether these match or not depends on the value of the parameter $q$. In order to see how this must be fixed, we need to discuss the global structure of the solution.

Let us first observe that by setting the rotational parameter $\alpha =0$, the $SU(2)\times U(1)$ symmetry of~\eqref{CGLP metricA}, \eqref{gaugefieldCGLP-A} is enhanced to $SO(4)$. This solution was originally found in~\cite{London:1995ib} and contains a naked singularity for any value of $q\neq 0$. So while the $\alpha = 0$ limit provides the natural symmetries to describe the vacuum of an SCFT on $S^3 \times \mathbb R$, it yields a solution that for any $q\neq 0$ is pathological, at least in supergravity. In appendix~\ref{SO4appendix} we prove that there are no other supersymmetric solutions with $SO(4)\times \mathbb R$ symmetry within minimal gauged supergravity. 

It was shown in~\cite{Cvetic:2005zi} that the two-parameter family of solutions given by \eqref{CGLP metricA}, \eqref{gaugefieldCGLP-A} contains a regular topological soliton (while there are no black holes free of CTC's). This is obtained by tuning the rotational parameter $\alpha$ to the critical value 
\be\label{criticalalpha}
\alpha^2 \, =\, q^3 \ .
\ee
Then the metric \eqref{CGLP metricA} has no horizon, is free of CTC's, and extends from $r=0$ to $\infty$. In addition, for the $r,\psi$ part of the metric to avoid a conical singularity while it shrinks as $r\to 0$, one has to impose
\be\label{regularq}
q \ = \ \frac{1}{9g^2}\ .
\ee
In this way one obtains a spin$^c$ manifold with topology $\mathbb R \times (\mathcal O(-1)\to S^2)$, where the first factor is the time direction, and the second has the topology of Taub-Bolt space~\cite{Cvetic:2005zi}. Since $\frac{\sqrt 3}{2}gA$ is a connection on a spin$^c$ bundle, as it can be seen from~\eqref{KillingSpEqDirac}, one must also check the quantisation condition for the flux threading the two-cycle at $r=0$. This reads
\be
\frac{1}{2\pi}\frac{\sqrt 3 }{2}g\int_{S^2} F  \  \in\   \mathbb Z  +  \frac{1}{2},
\ee
where the quantisation in half-integer units arises because the manifold is spin$^c$ rather than spin.
One can check that
\be\label{integralbolt}
\frac{1}{2\pi}\frac{\sqrt 3 }{2}g\int_{S^2} F \ = \ \frac{3}{2}  g\,q^{1/2}\ = \ \frac 12\ ,
\ee
hence the condition is satisfied.

We can then proceed to plug \eqref{regularq} into \eqref{Qcharge}. This gives 
\be
E \ =\ -\frac{\sqrt 3}{2} Q \ =\  \frac{2}{3}\,\frac{\pi^2}{g^2\kappa_5^2} \ ,
\ee
which is different from~\eqref{ExpectedCharges}. In field theory units, this reads $\langle R\rangle = \tfrac{4}{9}{\bf a}\neq \tfrac{16}{27}{\bf a}$, where the latter is the vev of the $R$-charge in a supersymmetric vacuum \cite{Assel:2015nca}  (recall footnote \ref{footy}). We conclude that although this 1/2 BPS topological soliton is smooth and  seemingly fullfills the requirements imposed by the field theory superalgebra, it is not dual to the vacuum state of an SCFT on the $S^3 \times \mathbb R$ background.  Below we will give further evidence that this  solution cannot describe  the supersymmetric 
vacuum state of a generic SCFT on $S^3\times \mathbb R$.

\subsection{Remarks on 1/2 BPS topological solitons}
\label{12remarks}

Firstly, we note that the non-trivial topology of the solution entails an obstruction to its embedding into string theory, precisely analogous to the situation of the ``bolt solutions'' found in 
\cite{Martelli:2012sz}.  For instance,  although the solution cannot be uplifted to type IIB supergravity on $S^5$~\cite{Cvetic:2005zi},
  there is a viable embedding if the orbifold  $S^5/\mathbb{Z}_3$ is chosen instead.  We discuss this issue in some detail in appendix~\ref{app:uplift}, where we also allow for a more general Lens space $S^3/{\mathbb{Z}_p}$ topology for the spatial part of the boundary geometry. 

Further information comes from studying regularity of the graviphoton $A$ in~\eqref{gaugefieldCGLP-A}. It was noted in~\cite{Cvetic:2005zi} that this is not well-defined as $r\to0$. Indeed, although $F_{\mu\nu}F^{\mu\nu}$ remains finite, $A_\mu A^\mu$ diverges as
\be
A_\mu A^\mu  \ = \ \frac{q}{r^2 }  + \mathcal O (1) ~ ,
\label{asquared}
\ee
where we have used the critical value $\alpha^2 = q^3$.
In order to cure this, one can introduce two new gauge potentials, $A'$ and $A''$, the first being well-defined around $r=0,$ $\theta = 0$, and the second being well-defined around $r=0$, $\theta = \pi$. These are related to the original $A$ by the gauge transformations\footnote{These gauge shifts have an opposite sign compared to those appearing in eq.~(3.36) of~\cite{Cvetic:2005zi}. To see this, one has to recall that $\alpha =q^{3/2}$ and take into account the different normalisation $A_{\rm here} = -\sqrt 3 A_{\rm there}$.
}
\bea\label{gaugeshifts}
A \ \longrightarrow \ A' \ = \ A + \frac{\sqrt3}{2}q^{1/2}(\diff \psi + \diff \phi)\ , \nn \\ [2mm]
A \ \longrightarrow \ A'' \ = \ A + \frac{\sqrt 3}{2}q^{1/2}(\diff \psi - \diff \phi) \ .
\eea
It was claimed in~\cite{Cvetic:2005zi} that the new gauge fields are not well-defined near to $r = \infty$ due to a singular term at order $\mathcal O(1/r^2)$, and for this reason a third gauge patch was introduced. However, we obtain a behavior different from the one displayed in eq.\ (3.37) of~\cite{Cvetic:2005zi}. We find
\bea
g^{\mu\nu} A'_\mu A'_\nu &=&  \frac{6q}{(r^2+q)(1+\cos\theta)} + \ldots \ ,\nn \\ [2mm]
g^{\mu\nu} A''_\mu A''_\nu &=& \frac{6q}{(r^2+q)(1-\cos\theta)} + \ldots \ ,
\eea
where the ellipsis denote a regular function of $r$ only.
The expressions are regular in the respective gauge patches. Hence it is not necessary to introduce a third gauge patch. Extending to infinity the two gauge patches introduced above, we obtain the boundary values
\bea
A'_\infty \ = \ c\, \diff t + \frac{\sqrt3}{2}q^{1/2}(\diff \psi + \diff \phi) \qquad\qquad \textrm{near}\ \theta =0\ , \nn \\ [2mm]
A''_\infty \ = \ c\, \diff t + \frac{\sqrt3}{2}q^{1/2}(\diff \psi - \diff \phi) \qquad\qquad \textrm{near}\ \theta =\pi \ ,
\eea
where $A_\infty$ is the graviphoton evaluated at $r=\infty$. 

Let us close this section with some comments on the interpretation of these flat fields in the putative field theory duals. $A_\infty$ is related to the background gauge field $A^{\rm cs}$ coupling canonically to the $R$-current of the dual SCFT by the conversion factor
$A^{\rm cs} = \frac{\sqrt 3}{2}gA_\infty$. Therefore, also using $q^{1/2} = \frac{1}{3g}$ and $c=-\frac{1}{\sqrt 3}$,  $A^{\rm cs}$ reads
\bea
A^{\rm cs} \ = \ -\frac{g}{2}\, \diff t + \frac{1}{4}(\diff \psi + \diff \phi) \qquad\qquad \textrm{near }\theta =0\ , \nn \\ [2mm]
A^{\rm cs} \ = \ -\frac{g}{2}\, \diff t + \frac{1}{4}(\diff \psi - \diff \phi) \qquad\qquad \textrm{near }\theta =\pi \ .
\eea
We see that in passing from the patch including $\theta = 0$ to the one including $\theta = \pi$, the gauge transformation $A^{\rm cs} \to A^{\rm cs} - \frac 12 \diff \phi$ is performed. Correspondingly, the dynamical fields in the dual SCFT acquire a phase $e^{-\frac{i}{2} q_R\phi}$, where $q_R$ is their $R$-charge. Since all bosonic {\it gauge invariant} operators in the SCFT should be well-defined in both patches as $\phi \to \phi + 2\pi$, we conclude that their $R$-charges must satisfy
 $q_R \in 2\mathbb Z$.\footnote{This condition is reminiscent of the quantisation of the $R$-charges which is imposed on supersymmetric field theories on $S^2 \times T^2$ by an $R$-symmetry monopole through $S^2$, see \emph{e.g.} \cite{Closset:2013sxa}. Note that if we impose the much more restrictive condition that the basic (scalar) fields of the gauge theory should have $R$-charge $q_R\in 2\mathbb{Z}$ then all known dual field theories would be ruled out because these have a superpotential with $R$-charge 2, implying all scalar fields in the theory have $R$-charges $< 2$.}
  We will make further comments in   appendix \ref{app:uplift}, where we will show that for a number of concrete examples this condition is automatically satisfied, after taking into account the constraints on the internal Sasaki-Einstein manifolds which follow from the conditions for uplifting the topological soliton to type IIB supergravity.

\subsection{Remarks on 1/4 BPS topological solitons}\label{sec:1/4BPStopsol}

We now consider the 1/4 BPS topological solitons mentioned in~\cite{Chong:2005hr}. We will show that under the assumption that the boundary has the topology of $S^3\times \mathbb R$, there are no regular topological solitons among the supersymmetric solutions of~\cite{Chong:2005hr} apart for the 1/2 BPS one discussed above.

Let us start from the boundary of the solution in~\cite{Chong:2005hr}. The boundary metric is obtained by sending $(gr)^2 \to + \infty$, and is given in \eqref{bdrymetrCCLP}. Requiring that this is (conformal to) the standard metric on $S^3 \times \mathbb R$ fixes the range of the coordinates as $\theta \in [0,\frac{\pi}{2}]$, $\phi \sim \phi + 2\pi$ and $\psi \sim \psi+2\pi$.\footnote{In the present subsection \ref{sec:1/4BPStopsol}, we drop  the label ``CCLP'' previously used to denote the coordinates of~\cite{Chong:2005hr}. One should recall anyway that these are not the same as the coordinates $\{r,\theta,\phi,\psi\}$ of~\cite{Cvetic:2005zi} appearing in subsections \ref{sec:compareCasimir}, \ref{12remarks}.} Moreover, 
requiring positivity of the spatial part of the boundary metric, {\it cf.} \eqref{bdrymetrCCLP}, we have that  the parameters $a$, $b$ should satisfy
\be\label{absmall}
|ag| <1\ ,\qquad  |bg|<1\ .
\ee
As discussed in~\cite{Chong:2005hr}, the condition that there is no horizon fixes the parameter $m$ as
\be
m \ =\ -(1 + a g) (1 + b g) (1 + ag + b g) (2a+ b +   a b g) (a + 2 b + a b g)\ .
\ee
Then the five-dimensional metric degenerates at
\be
r_0^2 \ =\ -(a+b+abg)^2 \ ,
\ee
(since the solution only depends on even powers of $r$, it can be continued to negative $r^2$). This is best seen by introducing a new radial coordinate
\be
(r')^2 \ =\ r^2 - r_0^2\ ,
\ee 
running from $r'=0$ to $\infty$, and making the change of angular coordinates
\bea
\phi &=& \phi'\ ,\nn \\ [2mm] 
\psi &=& \psi' +\frac{(1 - b g) (2 a + b + a b g)}{(1 - a g) (a + 2 b + a b g)}\, \phi'\ .
\label{changeangcoords}\eea
As $r'\to 0$, the orbit of $\frac{\partial}{\partial\phi'}$ shrinks to zero size, while the orbit generated by $\frac{\partial}{\partial\psi'}$ remains finite. Regularity requires that the orbit of $\frac{\partial}{\partial\phi'}$ is closed, hence the fraction in \eqref{changeangcoords} must be a \emph{rational number}. In this case, it follows that the angles $\phi', \psi'$ have the same periodicities as $\phi$, $\psi$, namely $2\pi$. 
Then one can see that absence of conical singularities in the $(r',\phi')$ plane as $r'\to 0$ requires~\cite{Chong:2005hr}
\be\label{noconical}
\left[\frac{(a+b+abg)(3 + 5ag + 5bg + 3abg^2)}{(1 - ag)(a + 2b + abg)}\right]^2 \ = \  1\ .
\ee

However, this is not the only condition needed for regularity. 
Let us consider the bolt surface at $r'=0$ and $t = {\rm const}$, parameterised by $\theta,\psi'$. One can see that as $\theta \to 0$, the leading terms of the metric on this surface are
\be
\diff s^2_{\rm bolt} \ = \  \frac{b(2a +b+abg)}{ag-1}\,\diff\theta^2 + \frac{b (a g-1) (a + 2 b + a b g)^2 }{(1 - b g)^2 (2a + b + a b g)}\,\theta^2\,(\diff \psi')^2 + \ldots\ . 
\ee
Therefore, the bolt has a conical singularity at $\theta = 0$ unless the additional condition
\be\label{noconicalBolt}
\left[\frac{(1-ag) (a + 2 b + a b g) }{(1 - b g) (2a + b + a b g)}\right]^2 \ = \ 1\ 
\ee
is satisfied. Notice that this implies $\psi = \psi' \pm  \phi'$ , which is consistent with the assumption we made that the two angular variables be related by a linear rational transformation in (\ref{changeangcoords}).
We have checked that the behaviour close to $\theta = \pi/2$ is always smooth instead.

The solutions to the regularity conditions \eqref{noconical}, \eqref{noconicalBolt} that also satisfy \eqref{absmall} are
\be\label{topsolCGLP-A}
a \ =\ - b\ =\ \pm \frac{1}{3g}
\ee
(corresponding to  $\psi = \psi' -  \phi'$ in   (\ref{changeangcoords})) and
\be\label{topsolCGLP-B}
a \ =\ b \ =\ \frac{-4+\sqrt{13}}{3g}
\ee
(corresponding to  $\psi = \psi' + \phi'$ in   (\ref{changeangcoords})).
Since $a=\pm b$, the corresponding solutions have $SU(2)\times U(1)$ invariance and were discussed in~\cite{Cvetic:2005zi}. The solution following from~\eqref{topsolCGLP-A} corresponds to the 1/2 BPS topological soliton already discussed above, contained in the ``case A'' of \cite[section$\:$3.3]{Cvetic:2005zi}. Case~\eqref{topsolCGLP-B} is contained in the ``case B'' solution of~\cite[section$\:$3.4]{Cvetic:2005zi}. In both cases, the metric on the bolt is the one of a round two-sphere.

Closer inspection reveals that only \eqref{topsolCGLP-A} is acceptable. Indeed, an additional constraint on the parameters comes from considering the $g_{\theta\theta}$ component of the five-dimensional metric in~\cite{Chong:2005hr}, that reads
\be
g_{\theta\theta} \ = \ \frac{(r')^2 - (a+b+abg)^2 + a^2\cos^2\theta + b^2\sin^2\theta}{1-a^2g^2\cos^2\theta-b^2g^2\sin^2\theta}\ .
\ee
This should remain positive all the way from $r'=\infty$ down to $r'=0$.
It is easy to check that while~\eqref{topsolCGLP-A} does satisfy the condition, \eqref{topsolCGLP-B} does not, implying that the signature of the metric changes while one moves towards small $r'$.

Therefore we conclude that among the supersymmetric solutions of~\cite{Chong:2005hr}, the only one corresponding to a completely regular topological soliton with an $S^3\times \mathbb R$ boundary is the 1/2 BPS soliton of~\cite{Cvetic:2005zi} that we considered above.

\section{Conclusions}\label{sec:conclusions}

In this paper, we studied supersymmetric solutions to minimal gauged supergravity in five dimensions, building on the approach of~\cite{Gauntlett:2003fk}. We derived a general expression for the sixth-order constraint that must be satisfied by the K\"ahler base metric in the timelike class of~\cite{Gauntlett:2003fk}, {\it cf.}~\eqref{constraintfinal}. We then considered a general ansatz comprising an orthotoric K\"ahler base, for which the constraint reduced to a single sixth-order equation for two functions, each of one variable, {\it cf.}~(\ref{orthomaster}). We succeded in finding an analytic solution to this equation, yielding a family of  AlAdS solutions with five non-trivial parameters. We showed that after setting two of the parameters to zero, this reproduces the solution of~\cite{Chong:2005hr} and hence encompasses (taking into account scaling limits) all AAdS solutions of the timelike class that are known within minimal gauged supergravity. This highlights the role of orthotoric K\"ahler metrics in providing supersymmetric solutions to five-dimensional gauged supergravity. For general values of the parameters, we obtained an AlAdS generalisation of the solutions of~\cite{Chong:2005hr}, of the type previously presented in \cite{Gauntlett:2003fk,Gauntlett:2004cm,Figueras:2006xx} in more restricted setups. There exists a further generalisation by an arbitrary anti-holomorphic function~\cite{Gauntlett:2003fk};  it would be interesting to study regularity and global properties of these AlAdS solutions.  

It would also be interesting to investigate further the existence of solutions to our  ``master equation'' (\ref{orthomaster}), perhaps aided by numerical analysis. 
In particular, our orthotoric setup could be used  as  the starting point for constructing a supersymmetric AlAdS solution dual to SCFT's on a squashed $S^3\times \mathbb R$ background, where the squashing of the three-sphere preserves just $U(1)\times U(1)$ symmetry. This would generalise the $SU(2)\times U(1)$ invariant solution of~\cite{Cassani:2014zwa}.

Finally, we have discussed the possible relevance of the solutions above to account for the non-vanishing supersymmetric vacuum energy and $R$-charge of a four-dimensional $\mathcal N=1$ SCFT defined on the cylinder $S^3\times \mathbb{R}$. The most obvious candidate for the gravity dual to the vacuum of an SCFT on $S^3\times \mathbb{R}$ is AdS$_5$ in global coordinates; however this comes with a vanishing $R$-charge. In appendix \ref{SO4appendix} we  have performed a complete analysis of supersymmetric solutions with $SO(4)\times \mathbb{R}$ symmetry, proving that there exists a unique singular solution, where the charge is an arbitrary parameter~\cite{London:1995ib}. Imposing regularity of the solution together with some basic requirements from
the supersymmetry algebra led us to focus on the 1/2 BPS smooth
topological soliton of~\cite{Cvetic:2005zi}. A direct evaluation of the energy and electric charge however showed that these do not match the SCFT vacuum expectation values.

We cannot exclude that there exist other solutions, possibly within our orthotoric
ansatz, or perhaps in the null class of \cite{Gauntlett:2003fk}, that match the supersymmetric Casimir energy of a four-dimensional $\mathcal N=1$ SCFT defined on the cylinder $S^3\times \mathbb{R}$.
It  would also be worth revisiting the evaluation of the charges of empty AdS space, and see if suitable boundary terms can shift the values of both the energy and electric charge, in a way compatible with supersymmetry.

Although we reached a negative conclusion about the relevance of the 1/2 BPS topological soliton to provide the gravity dual of the vacuum of a generic SCFT, 
our analysis clarified its properties, and may be useful for finding a holographic interpretation of this solution.   In appendix \ref{app:uplift} we showed that the embedding of this solution
into string theory includes simple internal Sasaki-Einstein manifolds such as $S^5$ and $T^{1,1}$. In particular, it should be possible to match the supergravity solution 
to a field theory calculation in the context of well-known theories such as ${\cal N}=4$ super Yang-Mills  placed on $S^3/{\mathbb{Z}_{3m}}\times \mathbb{R}$ and the Klebanov-Witten theory placed on $S^3/{\mathbb{Z}_{2m}} \times \mathbb{R}$.

\subsection*{Acknowledgments}

We would like to thank Zohar Komargodski and Alberto Zaffaroni for discussions. D.C. is supported by an European Commission Marie Curie Fellowship under the contract PIEF-GA-2013-627243. J.L. and D.M. acknowledge support by the ERC Starting Grant N. 304806, ``The Gauge/Gravity Duality and Geometry in String Theory.''

\appendix

\section{$SO(4)$-symmetric solutions}\label{SO4appendix}

In this appendix, we present an analysis of solutions to minimal gauged supergravity possessing $SO(4) \times \mathbb R$ symmetry. In particular, we prove that the only supersymmetry-preserving solution of this type is the singular one found long ago in~\cite{London:1995ib}. To our knowledge, a proof of uniqueness had not appeared in the literature before.

This appendix is somewhat independent of the rest of the paper, and the notation adopted here is not necessarily related to that.

The most general ansatz for a metric and a gauge field with $SO(4) \times \mathbb R$ symmetry is
\bea
\diff s^2 &=& - U(r) \dd t^2 + W(r) \dd r^2 +  2 X (r) \dd t \, \dd r  + Y(r) \dd \Omega_3^2 \ ,\\ [2mm]
A  &=& A_t (r) \dd t~,\label{gaugeansatz}
\eea
where $\diff \Omega_3^2$ is the metric on the round $S^3$ of unit radius,
\bea
\dd \Omega_3^2 & = & \frac{1}{4} \left( \sigma_1^2+\sigma_2^2+\sigma_3^2 \right)  ~ ,\nn\\
\sigma_1 &=& - \sin \psi \, \dd \theta  + \cos \psi \sin \theta \, \dd \phi ~ ,\nn\\
\sigma_2 &=& \cos \psi \, \dd \theta + \sin \psi \sin \theta \, \dd \phi ~ ,  \nn\\
\sigma_3 &=& \dd \psi + \cos \theta \, \dd \phi ~ .  \label{invariantforms}
\eea
The crossed term $X(r)\diff t\diff r$ in the metric can be removed by changing the $t$ coordinate, so we continue assuming $X(r)=0$.
We will make use of the frame
\be
e^0 \ =  \  \sqrt{U}\, \dd t ~ , \qquad e^{1,2,3} \ = \ \frac{1}{2} \sqrt Y  \sigma_{1,2,3} ~ ,\qquad e^4 \ = \ \sqrt W \, \dd r ~.    \label{frame}
\ee

\subsubsection*{Equations of motion}

We proceed by first solving the equations of motion and then examining the additional constraint imposed by supersymmetry. The action and equations of motion are given by equations \eqref{5Daction} and \eqref{5Deom}. With the ansatz \eqref{gaugeansatz}, the Maxwell equation is
\be
 0 \ = \ \nabla_\nu F^{\nu \mu } \qquad  \Leftrightarrow \qquad  0 \ = \ A_t'' + \frac12 A_t' \ \left( \log \frac{Y^3}{UW} \right)' ~ .
\ee
This can be integrated to
\be
A_t ' \ = \ c_1  \sqrt{\frac{U W }{Y^3}} ~ ,   \label{gaugefieldsolved}
\ee
with $c_1$ a constant of integration. 
The Einstein equations read (using frame indices)
\bea
R_{00} & = & -R_{44} \ = \  4 g^2 + \frac{(A_t')^2}{3UW}  ~ ,\nn\\
R_{11} & = & R_{22} \ = \ R_{33} \ = \  -4 g^2  +  \frac{(A_t')^2}{6UW}  ~ ,  \label{Einstein}
\eea
where the Ricci tensor components are 
\bea
R_{00} & =& \frac{U''}{2 U W}-\frac{U' W'}{4 U W^2}+\frac{3 U' Y'}{4 U W Y}-\frac{U'^2}{4 U^2 W} ~ ,\nn\\
R_{11} & =& R_{22} \ = \ R_{33} \ = \  -\frac{U' Y'}{4 U W Y}+\frac{W' Y'}{4 W^2 Y}-\frac{Y''}{2 W Y}-\frac{Y'^2}{4 W Y^2}+\frac{2}{Y} ~ , \nn\\
R_{44}  & =&-\frac{U''}{2 U W}+\frac{U' W'}{4 U W^2}+\frac{U'^2}{4 U^2 W}+\frac{3 W' Y'}{4 W^2 Y}-\frac{3 Y''}{2 W Y}+\frac{3 Y'^2}{4 W Y^2}~ .
\eea
To solve these, let us define
\be
T(r) \ = \ U(r)W(r)Y(r) ~ .  
\label{tuwy}
\ee
Combining two of the Einstein equations yields,
\be
0 \ = \ R_{00} + R_{44} \ = \ \frac{3 U }{4 T^2} \left(T' Y'-2 T Y''\right) ~ ,
\ee
which can be integrated to
\be
T(r)  \ = \ c_2  \, Y'^{\,2}(r)  ~ ,
\label{tyeq}
\ee
with $c_2 \neq 0$ a constant of integration. Using this, the angular components of the Einstein equations can  be integrated, yielding
\be
U(r)\ =\  4 c_2  +  4 c_2 g^2 Y+ \frac{1}{Y} c_3+ \frac{c_1^2 c_2}{3 Y^2}   ~ ,   \label{U}
\ee
with a third constant of integration $c_3$. This solves all the equations of motion.

We can now use the freedom to redefine the radial coordinate to choose one of the functions. In particular, we can choose the function $W(r)$ so that 
$ W U  =  4 s^2$,
where we take $s>0$.  From (\ref{tuwy}) and (\ref{tyeq}) we then obtain
\bea
\left(\frac{\diff Y}{\diff r}\right)^2 & = & \frac{4s^2}{c_2} Y  \qquad \Rightarrow \qquad Y(r) \ =\ \frac{s^2}{c_2}r^2   \ ,
\eea
where we used the freedom to shift $r$ to set to zero an integration constant.  Finally, after performing the trivial redefinitions
$r^{\rm old} = \frac{\sqrt{c_2}}{s} r^{\rm new}~,$ $U^{\rm old} = 4 c_2 U^{\rm new}$, $t^{\rm new} = 2\sqrt{c_2} t^{\rm old}$, we arrive at the solution
\bea
\diff s^2 & = & - U(r) \dd t^2 + \frac{1}{U(r)} \dd r^2 +  r^2 \dd \Omega_3^2 ~ , \label{metricU}\\
  A  & = & \left( c_4 - \frac{c_1 }{2r^2}\right) \diff t~,
\eea
with 
\be
U(r) \ = \ 1 +  g^2 r^2 + \frac{c_3}{ 4c_2 r^2}+ \frac{c_1^2}{12 r^4}      ~,   \label{U(r)}
\ee
and $c_4$ another arbitrary constant. Hence, the solution depends on three constants: $c_1$, which is essentially the charge, the ratio $c_3/c_2$, and $c_4$ which is quite trivial but may play a role in global considerations.

\subsubsection*{Supersymmetry}

The integrability condition of the Killing spinor equation \eqref{KillingSpEqDirac} is
\bea\label{integrabilitycond}
0 \ = \ \mathcal I_{\mu\nu}\epsilon &\equiv& \frac 14 R_{\mu\nu\kappa\lambda} \gamma^{\kappa\lambda}\epsilon + \frac{i}{4\sqrt 3} \left( \gamma_{[\mu}{}^{\kappa\lambda} + 4 \gamma^\kappa \delta_{[\mu}^\lambda \right) \nabla_{\nu]} F_{\kappa\lambda} \epsilon \nn \\[2mm]
&& +\,  \frac{1}{48} \left(F_{\kappa\lambda}F^{\kappa\lambda}\gamma_{\mu\nu} + 4 F_{\kappa\lambda}F^\kappa{}_{[\mu} \gamma_{\nu]}{}^\lambda - 6 F_{\mu\kappa}F_{\nu\lambda} \gamma^{\kappa\lambda} + 4F_{\kappa\lambda}F_{\rho [\mu} \gamma_{\nu]}{}^{\kappa\lambda\rho}\right) \epsilon \nn\\ [2mm]
&& + \,\frac{ig}{4\sqrt 3} \left( F^{\kappa\lambda}\gamma_{\kappa\lambda\mu\nu} - 4 F_{\kappa[\mu}\gamma_{\nu]}{}^\kappa - 6F_{\mu\nu} \right)\epsilon + \frac{g^2}{2}\gamma_{\mu\nu}\epsilon \  ,
\eea
where we used  $[\nabla_\mu , \nabla_\nu ] \epsilon   = \frac14 R_{\mu\nu\kappa\lambda}   \gamma^{\kappa\lambda}   \epsilon$. A necessary condition for the solution to preserve supersymmetry is that
\be
\mathrm{det}_\mathrm{Cliff} \,\mathcal I_{\mu\nu} \  = \ 0~  \qquad \textrm{for all}\ \mu,\nu\ ,
\ee
where the determinant is taken over the spinor indices. This gives for the $SO(4) \times \mathbb R$ invariant solution (in flat indices $a,b$):
\be
\mathrm{det}_\mathrm{Cliff}\, \mathcal I_{ab} \  = \  \frac{9\left(16 c_1^2 c_2^2-3 c_3^2\right)^2}{24^4 c_2^4\, r^{16}} \left(
\begin{array}{ccccc}
 0 & 1 & 1 & 1 & 81 \\
 1 & 0 & 1 & 1 & 1 \\
 1 & 1 & 0 & 1 & 1 \\
 1 & 1 & 1 & 0 & 1 \\
 81 & 1 & 1 & 1 & 0 \\
\end{array}
\right)_{ab}  ~ .
\ee
Hence, the supersymmetry condition is
\be
\frac{c_3}{c_2}  \ = \ - \frac{4}{\sqrt{3}} c_1 ~ ,
\ee
where we fixed a sign without loss of generality. Plugging this back into \eqref{U(r)}, we have
\bea
U(r) & = &  \left(1 -  \frac{c_1}{2\sqrt{3} \, r^2} \right)^2 + g^2 r^2 ~.   \label{Uagain}
\eea
This recovers a solution first found in~\cite{London:1995ib}. It is also obtained from~\eqref{CGLP metricA}--\eqref{gaugefieldCGLP-A} by setting $\alpha = 0$ and changing the radial coordinate.

Therefore we conclude that \emph{in the context of minimal gauged supergravity, the most general supersymmetric solution possessing $SO(4)\times\mathbb R$ symmetry is the one-parameter family found in \cite{London:1995ib}.} This preserves four supercharges and has a naked singularity. It would be interesting to determine if this can be acceptable in a string theory framework.

\section{Uplifting topological solitons to type IIB}\label{app:uplift}
  
In this appendix we discuss the uplift to type IIB supergravity of the 1/2 BPS topological soliton of~\cite{Cvetic:2005zi}. Recall that this is obtained from the solution in \eqref{CGLP metricA}--\eqref{gaugefieldCGLP-A} by choosing the rotational parameter as $\alpha^2 = q^3$ and fixing the remaining parameter $q$ so that conical singularities are avoided.
Compared to section~\ref{sec:compareCasimir}, we will consider the slightly more general case where the spatial part of the boundary has the topology of $S^3/\mathbb Z_p$ rather than just $S^3$. Then the periodicity of $\psi$ is $\frac{4\pi}{p}$ and the condition \eqref{regularq} on $q$ becomes 
\be\label{pdependentq}
q  \ =\ \frac{p^2}{9g^2}\ .
\ee
The four-dimensional hypersurfaces at constant $t$ have the topology of ${\cal O}(-p)\to S^2$. These manifolds are spin for $p$ even, while they are spin$^c$ for $p$ odd.

Locally, all solutions to five-dimensional minimal gauged supergravity can be embedded into type IIB supergravity on a Sasaki-Einstein five-manifold~\cite{Buchel:2006gb}. However, when the external spacetime has non-trivial topology one may encounter global obstructions. 
In particular, it was pointed out in~\cite{Cvetic:2005zi} that the 1/2 BPS topological soliton cannot be uplifted when the internal manifold is $S^5$. Here we identify the Sasaki-Einstein manifolds that make the uplift of that solution viable. 

The truncation ansatz for the ten-dimensional metric reads~\cite{Buchel:2006gb}
 \be\label{10Dmetric}
 \diff s^2_{10} \ =\  g_{\mu\nu}\diff x^\mu \diff x^\nu +\frac{1}{g^2}\left(\diff s^2(M) + \frac{1}{9}(\diff\zeta + 3\sigma - \sqrt 3 g A_\mu \diff x^\mu)^2 \right) 
 \ee
Following the presentation in \emph{e.g.}~\cite{Martelli:2004wu}, the metric 
 \be
\diff s^2 (SE) \ =\  \diff s^2(M) + \left(\frac{1}{3}\diff\zeta + \sigma \right)^2 
 \ee 
  is Sasaki-Einstein, where $\diff s^2(M)$ is an a priori local K\"ahler-Einstein metric, with K\"ahler two-form $J=\frac{1}{2} \diff\sigma$. The contact one-form is $\frac{1}{3}\diff\zeta + \sigma $
and the dual Reeb vector field is  $ 3 \frac{\partial}{\partial\zeta}\,.$ The graviphoton $A$ gauges the space-time dependent reparameterisations of $\zeta$.
  
 For eq.~\eqref{10Dmetric} to provide a good ten-dimensional metric, the one-form
 \be\label{10Doneform}
 \frac{1}{3}\diff\zeta + \sigma - \frac{g}{\sqrt 3}A
 \ee 
must be globally defined, and this imposes a constraint on the choice of Sasaki-Einstein manifold.
In particular, as discussed in \cite{Martelli:2012sz} in a closely related scenario, the one-form \eqref{10Doneform} can be globally defined only if $\zeta$ is periodically identified, thus one can never uplift to irregular Sasaki-Einstein manifolds. 
  
  Let us then  assume for simplicity to have a regular Sasaki-Einstein manifold,  that is a circle bundle over a Fano K\"ahler-Einstein manifold $M$, with Fano index $I(M)$. For example, for $M=\C P^2$ the Fano index is $I (\C P^2) =3$, while for $M=\C P^1\times \C P^1$ it is $I (\C P^1\times \C P^1)=2$ (see \emph{e.g.} \cite{Gauntlett:2006vf}).  Then, for any integer $k$ that divides $I$, the period of $\zeta$ can be taken $2\pi I/k$, with the Sasaki-Einstein five-manifold being simply connected if and only if $k=1$.  
Thus,   for $M=\C P^2$,  taking $\zeta$  to have period $6\pi$ gives $S^5$, while for   $M=\C P^1\times \C P^1$, a $\zeta$ with period $4\pi$ gives $T^{1,1}$.  If $k$ divides $I$ but is larger than one, a $\zeta$ with period $2\pi I/k$ yields a Sasaki-Einstein manifold that is still regular, albeit not simply connected. Defining $\tilde \zeta=\frac{k}{I}\zeta$, so that $\tilde \zeta$ has canonical period $2\pi $, we see that for the ten-dimensional metric~\eqref{10Dmetric} to be globally defined, the term 
  \be
  \diff \tilde\zeta - \frac{\sqrt 3 gk}{I} A
  \label{bonanza}
  \ee
  must be a \emph{bona fide} connection on a circle bundle, implying the quantisation condition
  \be
  \frac{\sqrt 3gk}{I}\int_{S^2} \frac{F}{2\pi} \  \in\ \mathbb{Z}~.
  \ee
 Using the computation in \eqref{integralbolt} with $q$ chosen as in \eqref{pdependentq}, we obtain
 \be
\frac{k\,p}{I}\ \in\ \mathbb{\mathbb{Z}}\  .
\label{pkI}
 \ee 
This condition relates the topology of the boundary manifold $\mathbb R\times S^3/\mathbb{Z}_p$ to the topology of the internal manifold. 
 
Let us now provide some examples of choices that obey (\ref{pkI}), together with some brief comments on the field theory duals.  We begin considering the case $p=1$
 as in the main body of the paper. 
For $M=\C P^2$ the condition (\ref{pkI})  implies that  $k=3$. This means that the Sasaki-Einstein manifold is $S^5/\mathbb{Z}_3$, and we can put the dual field theory
 on $S^3\times \mathbb R$.  This a quiver gauge theory with three nodes and nine bi-fundamental fields, arising from D3 branes placed at the ${\cal O}(-3) \to \C P^2$ singularity (see \emph{e.g.} \cite{Feng:2002zw}).
   According to the discussion at the end of section \ref{12remarks}, the $R$-charges of gauge-invariant operators of this theory, that may be constructed as closed loops of bi-fundamental fields in the quiver, must be even integers. This is in fact automatic, since the shortest loops are superpotential terms, that have $R$-charge precisely equal to 2.   Similarly, 
 for $M=\C P^1\times \C P^1$ condition (\ref{pkI}) with $p=1$
 implies $k=2$. Then the Sasaki-Einstein manifold is $T^{1,1}/\mathbb{Z}_2$, and we can put the dual field theory on $S^3 \times \mathbb R$. This is a quiver with 
 four nodes. In this case there are two possibilities for the bi-fundamental fields and superpotential, known as ``toric phases'' related by Seiberg duality  
  \cite{Beasley:2001zp,Feng:2002zw}. Again, one can check that all loops in the quiver are superpotential terms, and therefore have $R$-charge 2. 
 One can  go through all remaining regular Sasaki-Einstein cases, where the base is a del Pezzo surface $M=dP_i$, with $3 \leq i\leq 9$, which all have Fano index $I(dP_i)=1$, implying $k=1$. 
In fact,  according to (\ref{pkI}) these theories can be placed on  $S^3/\mathbb{Z}_p\times \mathbb{R}$ for any $p \geq 1$. 
For general $p$, the constraint on $R$-charges of gauge invariant operators derived in section  \ref{12remarks} is that these must be $q_R \in \tfrac{2}{p}\,\mathbb{Z}$. We have verified that for all four toric phases of the quivers dual to 
the third del Pezzo surface $M=dP_3$, the shortest loops are again superpotential terms  \cite{Beasley:2001zp}, and therefore satisfy this condition (for any $p$). 
 
There is in fact a more geometric way of  understanding the restriction on the choice of internal manifold $Y_5$, that is  directly 
related to the  field theory dual description. 
 Assuming that it is a regular Sasaki-Einstein, $Y_5$ can be identified with  the unit circle bundle in $\mathcal{L}=\mathcal{K}^{k/I}$, where $\mathcal{K}$ denotes 
the canonical line bundle of the K\"ahler-Einstein manifold $M$. 
Then scalar BPS  operators  in the dual field theory are in 1-1 correspondence with  holomorphic functions
on the Calabi-Yau cone over $Y_5$. These correspond to holomorphic 
sections of  $\mathcal{L}^{-n}$, with $n\in \mathbb{N}$ a positive integer.
Converting into field theory background $R$-symmetry gauge field the connection term in (\ref{bonanza})
this reads $- \tfrac{2k}{I}A^\mathrm{cs}$, showing that  the $R$-charge of the holomorphic functions is given by $q_R = \frac{2k}{I}n$.

Let us conclude  illustrating these general comments in two concrete  examples with $I>1$ and $p>1$. In particular, take $M=\C P^2$, and consider placing the theory on $S^3/\mathbb{Z}_{3m}\times \mathbb{R}$, thus picking $p=3m$. Then (\ref{pkI}) can be solved for either  $k=1$ or $k=3$. Choosing $k=1$ we can consider the theory dual to $S^5$, namely ${\cal N}=4$ super Yang-Mills. The gauge invariant operators in this theory are constructed with the three adjoints as Tr$(\Phi_I^{n_I} \Phi_J^{n_J}\Phi_K^{n_K})$, $I,J,K=1,2,3$, and  have
 $R$-charge equal to $q_R=\tfrac{2}{3}(n_I+n_J+n_K) \in  \tfrac{2}{3}\mathbb{N}$.

Finally, take $M=\C P^1\times \C P^1$, and consider placing the theory in   $S^3/\mathbb{Z}_{2m}\times \mathbb{R}$, thus picking $p=2m$. Then (\ref{pkI}) can be solved for either  $k=1$ or $k=2$. Choosing $k=1$ we can consider the theory dual to $T^{1,1}$, namely the Klebanov-Witten theory. 
The gauge invariant operators in this theory are constructed as Tr$(A_IB_J)^{n}$, where $A_I,B_J$, with $I,J=1,2$ 
are bi-fundamentals with $R$-charge equal to $1/2$. Thus the gauge invariant operator have $R$-charge $q_R = n\in \mathbb{N}$. 

It would be straightforward to generalise these considerations to the class of quasi-regular Sasaki-Einstein manifolds, where the K\"ahler-Einstein base is an orbifold.


\end{document}